\documentclass[a4paper]{cas-dc}

\usepackage{multirow}
\usepackage{booktabs}
\usepackage{pdflscape}
\usepackage[inline]{enumitem}
\usepackage{acronym}
\usepackage[numbers]{natbib}

\usepackage{xcolor}
\usepackage{hyperref}

\acrodef{CF}{collaborative filtering}
\acrodef{CRS}{conversational recommender system}
\acrodef{DIN}{deep interest network}
\acrodef{FM}{factorization machine}
\acrodef{GBDT}{gradient boosting decision tree}
\acrodef{IR}{information retrieval}
\acrodef{LR}{logistic regression}
\acrodef{NCF}{neural collaborative filtering} 
\acrodef{NLP}{natural language processing}
\acrodef{RS}{recommender system}
\acrodef{RL}{reinforcement learning}
\acrodef{DRL}{deep reinforcement learning}
\acrodef{TDM}{tree-based deep model}
\acrodef{LDA}{latent dirichlet allocation}
\acrodef{HCI}{human-computer interaction}
\acrodef{MAB}{multi-armed bandit}
\acrodef{PMF}{probabilistic matrix factorization}
\acrodef{MF}{matrix factorization}
\acrodef{DQN}{deep Q-network}
\acrodef{DDPG}{deep deterministic policy gradient}
\acrodef{CRM}{conversational recommender model}
\acrodef{EAR}{Estimation-Action-Reflection}
\acrodef{VAE}{variational autoencoder}
\acrodef{CPR}{conversational path reasoning}
\acrodef{GRU}{gated recurrent unit}
\acrodef{BERT}{bidirectional encoder representations from transformers}
\acrodef{MIM}{mutual information maximization}
\acrodef{KBRD}{Knowledge-based recommender dialog system}
\acrodef{AMT}{Amazon Mechanical Turk}
\acrodef{LSTM}{long short term memory}
\acrodef{GCN}{graph convolutional network}
\acrodef{MGCG}{multi-goal driven conversation generation}
\acrodef{RNN}{recurrent neural networks}

\makeatletter
\def\hlinew#1{%
  \noalign{\ifnum0=`}\fi\hrule \@height #1 \futurelet
   \reserved@a\@xhline}

\AtBeginDocument{%
  \providecommand\BibTeX{{%
    \normalfont B\kern-0.5em{\scshape i\kern-0.25em b}\kern-0.8em\TeX}}}

\newcommand{\myeqp}[1]{\hyperref[eq:#1]{Equation~\ref*{eq:#1}}}
\newcommand{\mysec}[1]{\hyperref[sec:#1]{Section~\ref*{sec:#1}}}
\newcommand{\mytable}[1]{\hyperref[tab:#1]{Table~\ref*{tab:#1}}}
\newcommand{\myfig}[1]{\hyperref[fig:#1]{Figure~\ref*{fig:#1}}}
\newcommand{\myappendix}[1]{\hyperref[appendix:#1]{Appendix~\ref*{appendix:#1}}}
\newcommand{\myalg}[1]{\hyperref[alg:#1]{Algorithm~\ref*{alg:#1}}}
\begin{document}
\title[mode = title]{Advances and Challenges in Conversational Recommender Systems: A Survey}
\shorttitle{Advances and Challenges in Conversational Recommender Systems: A Survey}


\author[ustc]{Chongming Gao}[orcid=0000-0002-5187-9196]
\ead{chongming.gao@gmail.com}
\ead[URL]{http://chongminggao.me}

\author[nus]{Wenqiang Lei}
\ead{wenqianglei@gmail.com}
\cormark[1]

\author[ustc]{Xiangnan He}
\ead{xiangnanhe@gmail.com}

\author[mdr1,mdr2]{Maarten de~Rijke}[orcid=0000-0002-1086-0202]
\ead{m.derijke@uva.nl}

\author[nus]{Tat-Seng Chua}
\ead{chuats@comp.nus.edu.sg}

\address[ustc]{University of Science and Technology of China}
\address[nus]{National University of Singapore}
\address[mdr1]{University of Amsterdam, Amsterdam, The Netherlands}
\address[mdr2]{Ahold Delhaize, Zaandam, The Netherlands}

\cortext[cor1]{Corresponding author}

\shortauthors{Chongming Gao et~al.}












\begin{abstract}
  Recommender systems exploit interaction history to estimate user preference, having been heavily used in a wide range of industry applications. However, static recommendation models are difficult to answer two important questions well due to inherent shortcomings: 
  \begin{enumerate*}[label=(\alph*)]
  \item What exactly does a user like?
  \item Why does a user like an item? 
  \end{enumerate*} 
  The shortcomings are due to the way that static models learn user preference, i.e., without explicit instructions and active feedback from users. The recent rise of \acp{CRS} changes this situation fundamentally. In a CRS, users and the system can dynamically communicate through natural language interactions, which provide unprecedented opportunities to explicitly obtain the exact preference of users.
  
  Considerable efforts, spread across disparate settings and applications, have been put into developing \acp{CRS}. Existing models, technologies, and evaluation methods for \acp{CRS} are far from mature. In this paper, we provide a systematic review of the techniques used in current \acp{CRS}. We summarize the key challenges of developing \acp{CRS} in five directions: 
  \begin{enumerate*}[label=(\arabic*)]
  \item Question-based user preference elicitation.
  \item Multi-turn conversational recommendation strategies. 
  \item Dialogue understanding and generation. 
  \item Exploitation-exploration trade-offs. 
  \item Evaluation and user simulation. 
  \end{enumerate*}
  These research directions involve multiple research fields like \ac{IR}, \ac{NLP}, and \ac{HCI}.
  Based on these research directions, we discuss some future challenges and opportunities.
  We provide a road map for researchers from multiple communities to get started in this area. We hope this survey can help to identify and address challenges in \acp{CRS} and inspire future research.
\end{abstract}


\maketitle
\acresetall

\section{Introduction}
\label{sec:intro}
Recommender systems have become an indispensable tool for information seeking. Companies such as Amazon and Alibaba, in e-commerce,  Facebook and Wechat, in social networking, Instagram and Pinterest, in content sharing, and YouTube and Netflix, in multimedia services, all have the need to properly link items (e.g., products, posts, and movies) to users. An effective recommender system that is both accurate and timely can help users find the desired information and bring significant value to the business. Therefore, the development of recommendation techniques continues to attract academic and industrial attention.

Traditional recommender systems, which we call \emph{static recommendation models} in this survey, primarily predict a user's preference towards an item by analyzing past behaviors offline, e.g., click history, visit log, ratings on items. Early methods, such as \ac{CF}~\cite{sarwar2001item,schafer2007collaborative}, \ac{LR}~\cite{nelder1972generalized}, \ac{FM} \cite{rendle2010factorization}, and \ac{GBDT}~\cite{ke2017lightgbm}, have been intensively used in practical applications due to the efficiency and interpretability.  
Recently, more complicated but powerful neural networks have been developed, including Wide \& Deep \cite{cheng2016wide}, \ac{NCF}~\cite{he2017ncf}, \ac{DIN} \cite{zhou2018deep}, \ac{TDM} \cite{TDM}, and \acp{GCN} \cite{ying2018graph,pmlr-v97-wu19e,he2020lightgcn}.

\medskip\noindent\textbf{Inherent Disadvantages of Static Recommendations.} \phantom{word}
Static recommendation models are typically trained offline on historical behavior data, which are then used to serve users online~\cite{YoutubeDNN}. Despite their wide usage, they fail to answer two important questions:

\begin{enumerate}[nosep,leftmargin=*]
  \item \emph{What exactly does a user like?} 
  The learning process of static models is usually conducted on historical data, which may be sparse and noisy. Moreover, a basic assumption of static models is that all historical interactions represent user preference. Such a paradigm raises critical issues. First, users might not like the items they chose, as they may make wrong decisions~\cite{wang2020denoising,wang2020click}. Second, the preference of a user may drift over time, which means that a user's attitudes towards items may change, and capturing the drifted preference from past data is even harder~\citep{jagerman-when-2019}. In addition, for cold users who have few historical interactions, modeling their preferences from data is difficult~\citep{lee2019melu}. Sometimes, even the users themselves are not sure of what they want before being informed of the available options \cite{wang2013research}. In short, a static model can hardly capture the precise preference of a user.

  \item \emph{Why does a user like an item?} 
  Figuring out why a user likes an item is essential to improve recommender model mechanisms and thus increase their ability to capture user preference. 
  There are many factors affecting a user's decisions in real life~\citep{ma2019learning,cen2020kdd,gao2019bloma}. For example, a user might purchase a product because of curiosity or being influenced by others \cite{yu2019generating}. Or it may be the outcome of deliberate consideration. It is common that different users purchase the same product but their motivations are different. Thus, treating different users equally or treating different interactions by the same user equally, is not appropriate for a recommendation model. In reality, it is hard for a static model to disentangle different reasons behind a user's consumption behavior.
\end{enumerate} 

\noindent%
Even though much effort has been done to eliminate these problems, they make limited assumptions. For example, a common setting is to exploit a large amount of auxiliary data (e.g., social networks, knowledge graphs) to better interpret user intention~\cite{10.1145/2556270}. However, these additional data may also be incomplete and noisy in real applications. We believe the key difficulty stems from the inherent mechanism: the static mode of interaction modeling fundamentally limits the way in which user intention can be expressed, causing an asymmetric information barrier between users and machines.

\begin{figure}[pos=t]
\centering
\includegraphics[width=1\linewidth]{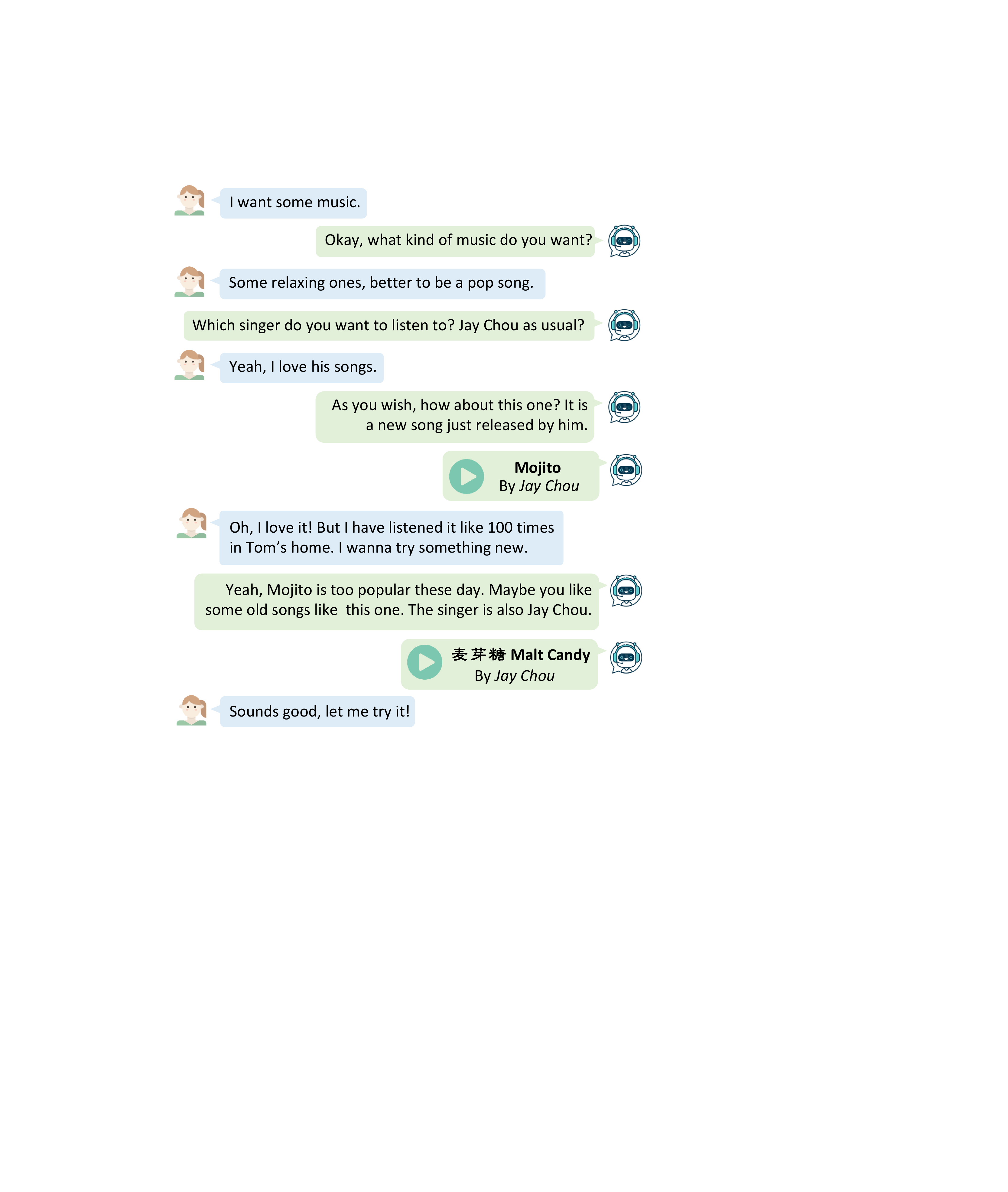}
\caption {A toy example of a \acl{CRS} in music recommendation.}
\label{fig:example}
\end{figure}

\medskip\noindent\textbf{Introduction of CRSs.}
The emergence of \acp{CRS} changes this situation in profound ways. There is no widely accepted definition of CRS. In this paper, we define a \ac{CRS} to be:
\begin{quote}
\emph{A recommendation system that can elicit the dynamic preferences of users and take actions based on their current needs through real-time multi-turn interactions.}
\end{quote}
Our definition highlights a property of \acp{CRS}: \emph{multi-turn interactions}. By a narrow definition, conversation means multi-turn dialogues in the form of written or spoken natural language; from a broader perspective, conversation means any form of interactions between users and systems, including written or spoken natural language, form fields, buttons, and even gestures \cite{Jannach2020ASO}. Conversational interaction is a natural solution to the long-standing asymmetry problem in information seeking. Through interactions, \acp{CRS} can easily elicit the current preference of a user and understand the motivations behind a consumption behavior.
\myfig{example} shows an example of a \ac{CRS} where a user resorts to the agent for music suggestions. Combining the user's previous preference (loving Jay Chou's songs) and the intention elicited through conversational interactions, the system can offer desired recommendations easily. Even if the produced recommendations do not satisfy the user, the system has chances to change recommendations based on user feedback.

\begin{figure}[pos=t]
\centering
\begin{tabular}{c}
\includegraphics[width=0.95\linewidth]{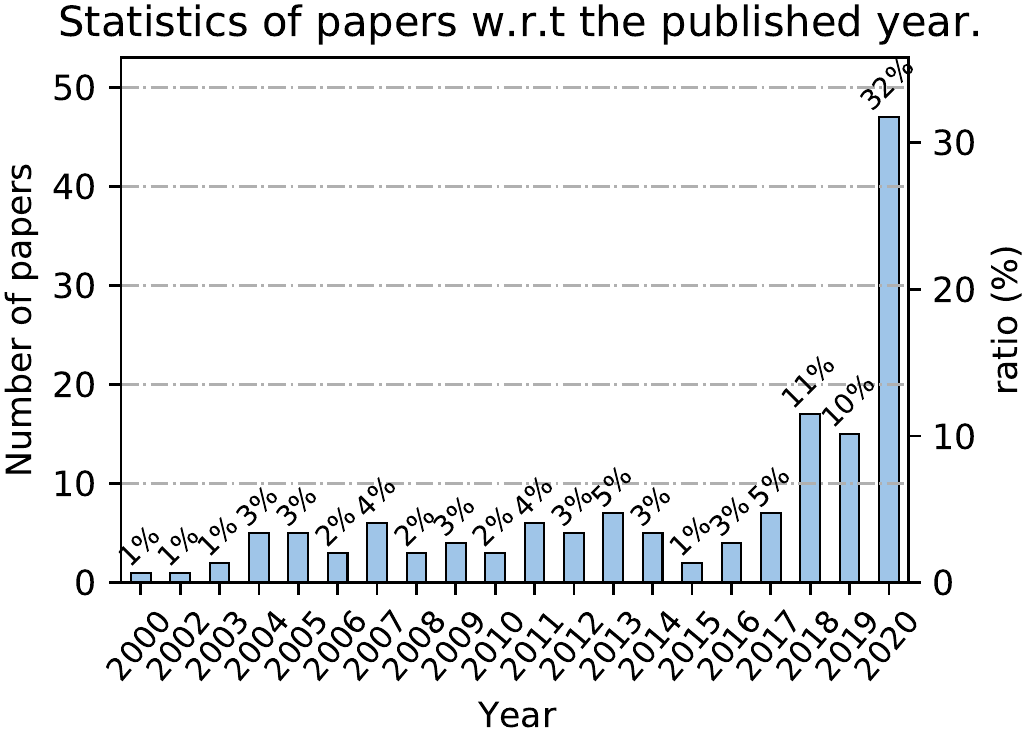}
\vspace{2mm}
\\\includegraphics[width=0.8\linewidth]{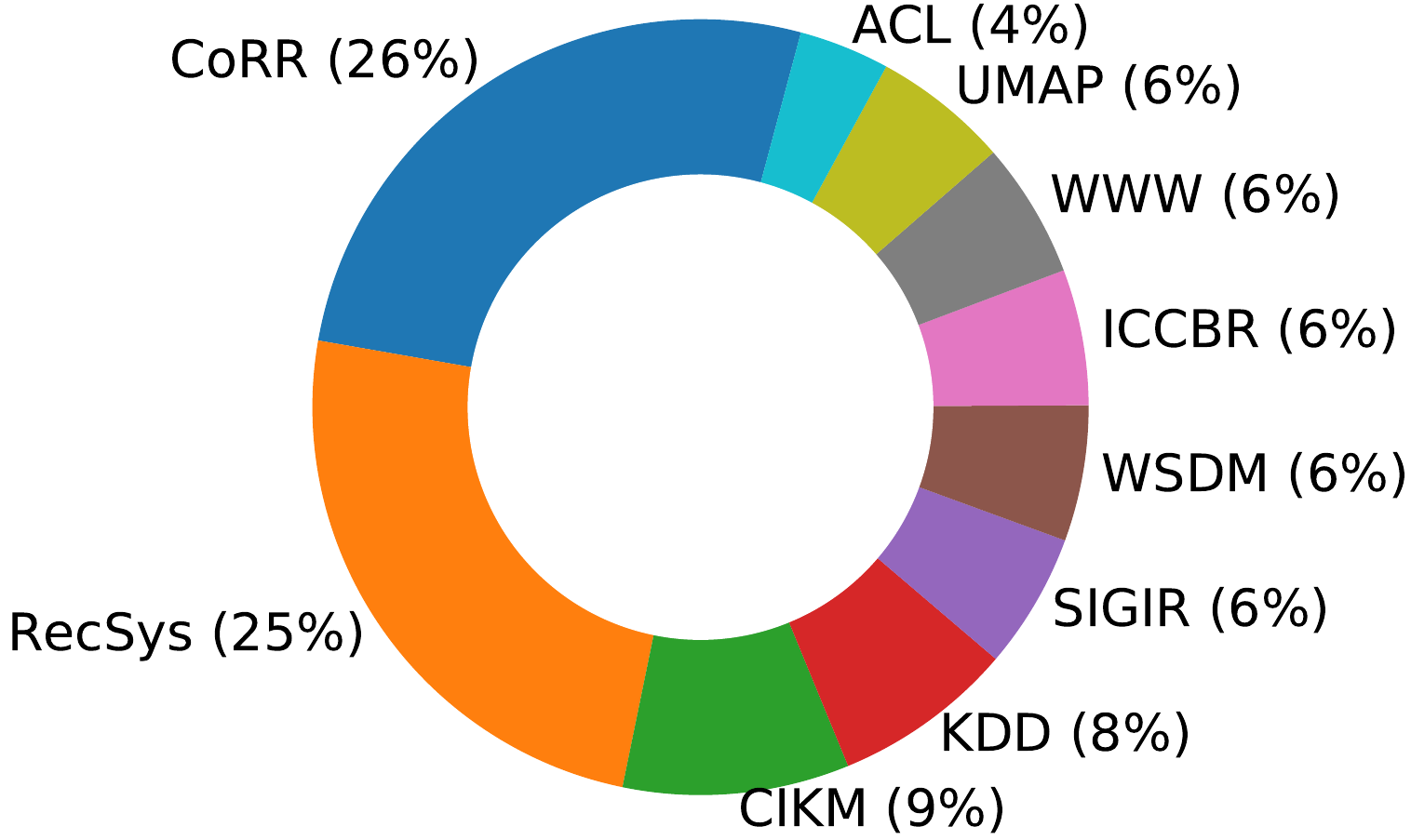}
\end{tabular}
\caption {Statistics of the publications related to \acp{CRS}, grouped by the publication year and venue. Only the top $10$ venues are used in the visualization.}
\label{fig:paper_statistics}
\end{figure}

Recently, attracted by the power of \acp{CRS}, many researchers have been on focusing on exploring this topic. These efforts are spread across a broad range of task formulation, in diverse settings and application scenarios. We collect the papers related to \acp{CRS} by searching for ``Conversation* Recommend*'' on DBLP\footnote{\url{https://dblp.org/search?q=conversation*\%20recommend*}} and visualize the statistics of them with regard to the published year and venue in \myfig{paper_statistics}. There are $148$ unique publications up to 2020, and we only visualize the top $10$ venues, which contain $53$ papers out of all $148$ papers at all $89$ venues. It is necessary to summarize these studies which put efforts into different aspects of \acp{CRS}.

\medskip\noindent\textbf{Connections with Interactive Recommendations.}
Since the born of recommender systems, researchers have realized the importance of the human-machine interaction. Some studies propose interactive recommender systems \cite{he2016interactive,wang2017factorization,chen2019large,zhou2020interactive} and critiquing-based recommender systems \cite{tou1982rabbit,tversky1993context,burke1997findme,smyth2003analysis,pu2004decision,chen2012critiquing,luo2020deep,luo2020latent}, which can be viewed as early forms of \acp{CRS} since they focus on improving the recommendation strategy online by leveraging real-time user feedback on previously recommended items. 

In the setting of interactive recommendations, each recommendation is followed by a feedback signal indicating whether and how much the user likes this recommendation. However, interactive recommendations suffer from low efficiency, as there are too many items. An intuitive solution is to leverage attribute information of items, which is self-explanatory for understanding users' intention and can quickly narrow down candidate items. The critiquing-based recommender system is such a solution that is designed to elicit users' feedback on certain attributes, rather than items. 
Critiquing is like a salesperson who collects user preference by asking questions proactively on item attributes. For example, when seeking mobile phones, a user may follow the hint of the system and provides feedback such as ``cheaper'' or ``longer battery life.'' Based on such feedback, the system will recommend more appropriate items; this procedure repeats several times until the user finds satisfactory items or gives up. The mechanism gives the system an improved ability to infer user preference and helps quickly narrow down recommendation candidates.

Though effective, existing interactive and critiquing methods have a limitation: the model makes a recommendation each time after receiving user feedback, which should be avoided as the recommendation should only be made when the confidence is high. This problem is solved in some \acp{CRS} by developing a conversation strategy determining when to ask and recommend \cite{lei20estimation,lei2020interactive}. Besides, the interactive and critiquing methods are constrained by their representation ability since users can only interact with the system through a few predefined options. The integration of a conversational module in \acp{CRS} allows for more flexible forms of interaction, e.g., in the form of tags \cite{christakopoulou2018q}, template utterances \cite{Sun:2018:CRS:3209978.3210002}, or free natural language \cite{nips18/DeepConv}. Undoubtedly, user intention can be more naturally expressed and comprehended through a conversational module.

\tabcolsep=1pt

\medskip\noindent\textbf{Connections with Other Conversational AI Systems.} \phantom{word}
Besides \acp{CRS}, there are other conversational AI systems, e.g., task-oriented dialogue systems \cite{chen2017survey,zhang2020recent,pei-2021-cooperative}, social chatbots \cite{ma2021one,li-2021-improving,wu2018deep}, conversational searching \cite{voskarides-2020-query,rosset2020leading,ren-2021-wizard}, and conversational question answering (QA) \cite{zhu2021retrieving}. The common point of them is to utilize natural language as a powerful tool to convey information and thus to provide a natural user interface. Though these research topics all possess the keyword ``conversation'', the central tasks are different. For example, while task-oriented dialogue systems aim to fulfill a certain task in human-machine dialogue, the concentration of effort is mainly on handling information in the textural language-based dialogue, e.g., natural language understanding (NLU), dialogue state tracking (DST), dialogue policy learning (DPL), and natural language generation (NLG) \cite{chen2017survey,zhang2020recent,gao2019neural}. In \acp{CRS}, however, the multi-turn conversation can be built on any form of interaction (e.g., form fields, buttons, and even gestures \cite{Jannach2020ASO}) instead of merely textual form. Because \acp{CRS} concentrate on recommendation logic, the textual dialogue is just one possible means to convey information, i.e., it is auxiliary, not necessary. 
Although there are some \acp{CRS} implemented as end-to-end dialogue systems \cite{nips18/DeepConv,chen-etal-2019-towards}, the human evaluation conducted by \citet{jannachend2020} suggests the performance is not ideal and more efforts should be put on improving both recommendation and language generation.

Other conversational AI systems can also be distinguished from \acp{CRS} by their specific scenarios. For instance, conversational searching focuses on analyzing the input query (in contrast to eliciting user preference in \acp{CRS}); conversational QA focuses on the single-turn question answering (in contrast to multi-turn interaction in \acp{CRS}). Therefore, it is essential to identify the central tasks and primary challenges in \acp{CRS} to help the beginner and future researchers set foot in this field and keep up with state-of-the-art technologies. 

\begin{figure*}[pos=t,align=\centering]
\centering
\includegraphics[width=0.85\linewidth]{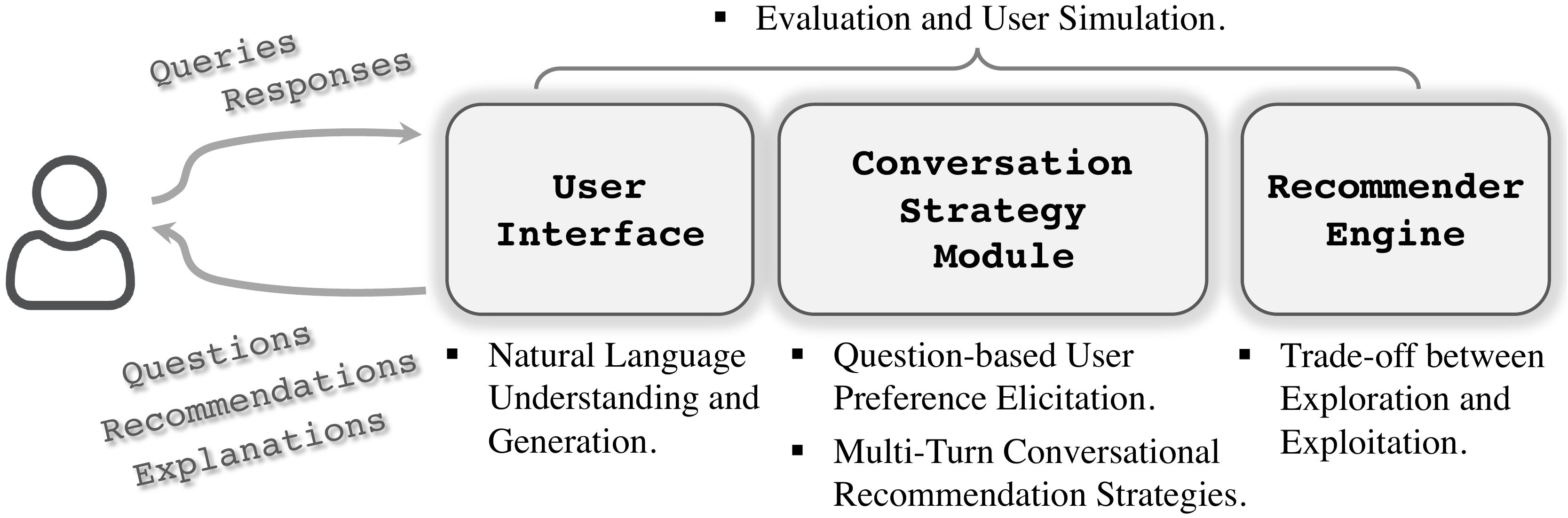}
\caption {Illustration of the general framework of \acp{CRS} and our identified five primary challenges on the three main components.}
\label{fig:framework}
\end{figure*}

\medskip\noindent\textbf{Focuses of This Survey.}
Although many studies have been done on \acp{CRS}, there is no uniform task formulation. In this survey, we present all \acp{CRS} as the general framework that consists of three decoupled components illustrated in \myfig{framework}.
Specifically, a \ac{CRS} is made of a user interface, a conversation strategy module, and a recommendation engine. The user interface serves as a translator between the user and machine; generally, it extracts information from raw utterances of the user and transforms the information into machine-understandable representation, and it generates meaningful responses to the user based on the conversation strategy. The conversation strategy module is the brain of the \ac{CRS} and coordinates the other two components; it decides the core logic of the \ac{CRS} such as eliciting user preference, maintaining multi-turn conversations, and leading new topics. The recommendation engine is responsible for modeling relationships among entities (e.g., the user-item interaction or item-item linkage), learning and recording user preference on items and attributes of items, retrieving the required information. 

There are many challenges in the three components, we summarize five main challenges as following.

\begin{itemize}[nosep,leftmargin=*]
    \item \emph{Question-based User Preference Elicitation.} 
    \acp{CRS} provide the opportunity to explicitly elicit user preference by asking questions. Two important questions are needed to be answered: 
    \begin{enumerate*}[label=(\arabic*)]
    \item What to ask?
    \item How to adjust the recommendations based on user response? 
    \end{enumerate*}
    The former focuses on constructing questions to elicit as much information as possible; the latter leverages the information in user response to make more appropriate recommendations.
    
    \item \emph{Multi-turn Conversational Recommendation Strategies.} \phantom{word} 
    The system needs to repeatedly interact with a user and adapts to the user's response dynamically in multiple turns. An effective strategy concerns when to ask questions and when to make recommendations, i.e., let the model choose between
    \begin{enumerate*}[label=(\arabic*)]
    \item continuing to ask questions so as to further reduce preference uncertainty, and 
    \item generating a recommendation based on estimation of current user preference.
    \end{enumerate*}
    Generally, the system should aim at a successful recommendation using the least number of turns, as users will lose their patience after too many turns~\citep{lei20estimation}. Furthermore, some sophisticated conversational strategies try to proactively lead dialogues \cite{wu-etal-2019-proactive,balaraman2020proactive}, which can introduce diverse topics and tasks in \acp{CRS}~\cite{liu2020ACL,zhou2020topicguided,lewis-etal-2017-deal,wang-etal-2019-persuasion}.
    
    \item \emph{Natural Language Understanding and Generation.}\\
    Communicating like a human being continues to be one of the hardest challenges in \acp{CRS}. 
    For understanding user interests and intentions, some \ac{CRS} methods define the model input as pre-defined tags that capture semantic information and user preferences \cite{christakopoulou2018q,lei20estimation,lei2020interactive,10.1145/3397271.3401180}. Some methods extract the semantic information from users' raw utterances via slot filling techniques and represent user intents in slot-value pairs \cite{zhang2018towards,Sun:2018:CRS:3209978.3210002,10.1145/3394592}. And for generating human-understand\-able responses, \acp{CRS} use many strategies such as directly providing a recommendation list \cite{10.1145/3397271.3401180,zhang2018towards}, incorporating recommended items in a rule-based natural language template \cite{Sun:2018:CRS:3209978.3210002,lei20estimation,lei2020interactive}. Moreover, some researchers propose the end-to-end framework to enable \acp{CRS} to precisely understand users' sentiment and intentions from the raw natural language and to generate readable, fluent, consistent, and meaningful natural language responses \cite{nips18/DeepConv,liu2020ACL,10.1145/3394592,chen-etal-2019-towards,zhou2020improving}. 
    
    \item \emph{Trade-offs between Exploration and Exploitation (E\&E).} 
    One problem of recommender systems is that each user can only interact with a few items out of the entire dataset. A large number of items that a user may be interested in will remain unseen by the user. For cold-start users (who have just joined the system and have zero or very few interactions), the problem is especially severe. Thanks to the interactive nature, \acp{CRS} can actively explore the unseen items to better capture the user preference. In this way, users can benefit from having chances to express their intentions and obtain better-personalized recommendations. However, the process of exploration comes at a price. As users only have limited time and energy to interact with the system, a failed exploration will waste time and lose the opportunity to make accurate recommendations. Moreover, exposing unrelated items hurts user preference, compared to exploiting the already captured preference by recommending the items of high confidence \cite{schnabel2018short,li2015toward,gilotte2018offline}.
    Therefore, pursuing E\&E trade-offs is a critical issue in \acp{CRS}.
    
    \item \emph{Evaluation and User Simulation.}
    Evaluation is an important topic. Unlike static recom\-men\-der models that are optimized on offline data, \acp{CRS} emphasize the user experience during dynamic interactions. Hence, we should not only consider the turn-level evaluation for both recommendation and response generation but also pay attention to the conversation-level evaluation. Besides, evaluating \acp{CRS} requires a large number of online user interactions, which are expensive to obtain \cite{li2015toward,jagerman-when-2019,huang-2020-keeping}. Practical solutions include: 
    \begin{enumerate*}[label=(\arabic*)]
    \item leveraging the off-policy evaluation which assesses the target policy using the logged data under the behavior policy \cite{gilotte2018offline,jagerman-when-2019}, and
    \item directly introducing user simulators to replace the true users in evaluation \cite{simulation2020kdd,sun-2021-simulating}.
    \end{enumerate*}
\end{itemize}

\noindent
The five challenges are allocated to the corresponding component as illustrated in \myfig{framework}, where trading off the E\&E balance is exclusive to the recommender engine; handling natural language understanding and generation is exclusive to the conversation module. The rest three challenges are related to both the components. We illustrate in \mytable{sum} the solutions of some classic \acp{CRS} that focus on these directions. Limited by space, we only give part of the classic studies here. We will further discuss existing solutions in the following sections.

\begin{table*}[pos=t]
\tabcolsep=5pt
\renewcommand\arraystretch{1.7}
\caption{Five primary challenges in \acp{CRS} and part of the classic methods that contribute to these challenges.}
\label{tab:sum}
\begin{tabular}{@{}lll@{}}
\toprule
\textbf{Primary Challenges in \acp{CRS}}                    & \textbf{Contributions of Existing Studies} & \textbf{Classic Publications}                                                                                                                                                        \\ \midrule
\multirow{2}{*}{Question-based User Preference Elicitation} & Asking about items                         & \cite{cikm13/ICF,christakopoulou2016towards,10.1145/3292500.3330991,zou2020neural,mangili2020bayesian,vendrov2020gradient,loepp2014choice}                                           \\ \cmidrule(l){2-3} 
                                                            & Asking about attributes                    & \cite{mangili2020bayesian,Wu:umap21,10.1145/3298689.3347009,christakopoulou2018q,zhang2018towards,Sun:2018:CRS:3209978.3210002,lei20estimation,lei2020interactive,zhou2020improving} \\ \midrule
\multirow{2}{*}{Multi-turn Conversational Strategies}       & Explicit  strategies                       & \cite{Sun:2018:CRS:3209978.3210002,zhang2018towards,lei20estimation,xu2021adapting}                                                                                                  \\ \cmidrule(l){2-3} 
                                                            & Leading diverse topics                     & \cite{liu2020ACL,zhou2020topicguided}                                                                                                                                                \\ \midrule
Language Understanding and Generation                       & End-to-end dialogue systems                & \cite{nips18/DeepConv,chen-etal-2019-towards,zhou2020improving,xu2020user,moon-2019-opendialkg}                                                                                      \\ \midrule
Exploration and Exploitation Trade-offs                     & Leveraging multi-armed bandits             & \cite{christakopoulou2016towards,zhang2020conversational,li2020seamlessly,10.1145/3292500.3330991}                                                                                   \\ \midrule
\multirow{2}{*}{Evaluation and User Simulation}             & Evaluation                                 & \cite{gilotte2018offline,huang-2020-keeping}                                                                                                                                         \\ \cmidrule(l){2-3} 
                                                            & User simulation                            & \cite{simulation2020kdd,sun-2021-simulating}                                                                                                                                         \\ \bottomrule
\end{tabular}
\end{table*}

\medskip\noindent\textbf{Differences with Existing Related Surveys.}
Recently, A number of related survey papers have been published. There are survey papers focusing on certain cutting-edge aspects in recommender systems, such as the bias issues and debiasing methods \cite{chen2020bias}, explainability/interpretability \cite{Zhang_2020}, evaluation issues \cite{silveira2019good}, and novel methods that leverage deep neural networks \cite{zhang2019deep,wu2021survey,wu2020graph}, knowledge graphs \cite{guo2020survey}, or reinforcement learning \cite{afsar2021reinforcement} to improve the ability of recommendation systems. Also, there are survey papers that summarize new frontiers in conversational AI systems, such as the advanced methods \cite{chen2017survey,zhang2020recent,gao2019neural} and the evaluation issues \cite{celikyilmaz2020evaluation,deriu2021survey} in dialogue systems. However, there is only one survey paper published in 2020 that focuses on \acp{CRS} \cite{Jannach2020ASO}. 

\citet{Jannach2020ASO}, for the first time, delved into different aspects of \acp{CRS} and made a comprehensive survey of \acp{CRS}. Specifically, they categorize existing \acp{CRS} in various dimensions, for instance, in terms of interaction modalities (e.g., buttons or written language), supported tasks (e.g., recommend or explain), or the knowledge \acp{CRS} use in the background (e.g., item-related information or dialogue corpora). Their survey provides a structured description of the \ac{CRS}. Therefore, the audience, after reading this survey, can answer \emph{what a \ac{CRS} is}, for example, what the input/output or the functions of a \ac{CRS} are. However, they may be still unsure about \emph{what the key challenges are}, or \emph{what to do next}. In our survey, we not only give the review of the current progress on \acp{CRS} including the existing assumptions and exploration but also refine the problems in state-of-the-art methods and summarize five challenges. We are trying to answer the three questions above, and we hope to provoke deeper thought and spark new ideas for the audience.


\medskip\noindent\textbf{Survey Organization.}
The remainder of this paper is organized as follows. In next several sections, we discuss the main challenges in \acp{CRS}. Specifically, in \mysec{question-driven}, we illustrate how \acp{CRS} can elicit user preferences by asking informative questions. In \mysec{multi-turn}, we describe the strategies in \acp{CRS} to interact with users in a multi-turn conversation. In \mysec{NLUNLG}, we point out the problems and provide solutions in dialogue understanding and generation for \acp{CRS}. In \mysec{EE}, we discuss how \acp{CRS} can balance the exploration-exploitation trade-off. In \mysec{evaluation}, we explore metrics and present techniques for evaluating \acp{CRS}. In \mysec{future-directions}, we envision some promising future research directions. And in \mysec{conclusion}, we conclude this survey.

\section{Question-based User Preference Elicitation}
\label{sec:question-driven}

A user looking for items with specific attributes may get assess to them by actively searching. For instance, a user may search ``iphone12 red 256gb'', where the key phrases ``red'' and ``256gb'' are the attributes of the item iPhone12. In this scenario, users construct a query themselves, and the performance relies on both the search engine and the user's expertise in constructing queries. Even though there are efforts on helping users complete queries by suggesting possible options based on what they entered \cite{ma2008learning,bar2011context,dehghani2017learning,cai-survey-2016}, users still need to figure out appropriate query candidates. Besides, searching in this way requires users to be familiar with each item they want, which is not true in practice. Recommender systems introduce users to the potential items that they may like. However, traditional recommender systems can only utilize the static historical records as the input, which results in the two main limitations mentioned in mysec{intro}.

Fortunately, \acp{CRS} can bridge the gap between the search engine and recommender system. Empowered by real-time interactions, \acp{CRS} can proactively consult users by asking questions. And with the feedback returned by users, \acp{CRS} can directly comprehend users' needs and attitudes towards certain attributes, hence making proper recommendations. Even if users are not satisfied with the recommended items, a \ac{CRS} has the opportunity to adjust its recommendations in the interaction process. 

Question-driven methods focus on the problem of \emph{what to ask} in conversations. Generally, there are two kinds of methods: 
\begin{enumerate*}[label=(\arabic*)]
\item asking about items \cite{cikm13/ICF,christakopoulou2016towards,sepliarskaia2018preference}, or 
\item asking about attributes/topics/categories of items \cite{lei20estimation,lei2020interactive}. 
\end{enumerate*}

\subsection{Asking about Items}
Early studies directly ask users for opinions about an item itself \cite{cikm13/ICF,wang2018online,christakopoulou2016towards,zou2020neural,vendrov2020gradient}. Unlike traditional recommender systems which need to estimate user preferences in advance, \acp{CRS} can construct and modify the user profile during the interaction process.

In traditional recommender system models, the recommended items are produced in a relatively stable way from all candidates. In the \ac{CRS} scenario, the recommended items should be updated after the system receives feedback from a user and it could be a complete change in order to adapt to the user's real-time preferences. Hence, instead of merely updating parameters of models online, some explicit rules or mechanisms are required. We introduce three methods that can elicit users' attitudes towards items and can quickly adjust recommendations. Most of these methods did not use natural language in their user interface, but it can easily integrate an natural language-based interface to make a \ac{CRS}.

\medskip\noindent\textbf{Choice-based Methods.}
The main idea of choice-based preference elicitation is to recurrently let users choose their preferred items or item sets from the current given options. The common strategies include
\begin{enumerate*}[label=(\arabic*)]
\item choosing an item from two given options \cite{sepliarskaia2018preference},
\item selecting an item from a list of given items \cite{jiang2014choice,graus2015improving,saavedra2016choice}, and
\item choosing a set of items from two given lists \cite{loepp2014choice}.
\end{enumerate*}
After the user chooses preferred items, the methods change the recommendations according to the user's choice. For example, \citet{loepp2014choice} use the \ac{MF} model \cite{bell2007modeling} to initialize the embedding vectors of users and items, then select two sets of items from the item embedding space as candidate sets and let a user choose one of the two sets. It is important to ensure that the two candidate sets are as different or distinguishable as possible. To achieve this, the authors adopt a factor-wise MF algorithm \cite{bell2007modeling}, which factorizes the user-item interaction matrix and obtains the embedding vectors one by one in decreasing order of explained variance. Hence, the factors, i.e., different dimensions of embedding vectors, are ordered by distinctiveness. Then, the authors iteratively select two item sets with only a single factor value varying. For example, if two factors represent the degree of \emph{Humor} and \emph{Action} of movies, respectively, then the two candidate sets are one set of movies with a high degree of \emph{Humor} and another with a low degree of \emph{Humor}, while the degree of \emph{Action} of the two sets is fixed to the average level.  When a user chooses one item set, the user's preference embedding vector is set to the average of the embedding vectors of the chosen items. The choice becomes harder as the interaction process continues. Users can choose to ignore the question, which means the users cannot tell the difference between the two item sets or they do not care about it. \citet{carenini2003towards} further explore other strategies to select query items, e.g., selecting the most popular or the most diverse items in terms of users' history.

\medskip\noindent\textbf{Bayesian Preference Elicitation.}
In addition, there are studies based on a probabilistic view of preference elicitation, which has been researched for a long time \cite{Chajewska1998UAI,boutilier2002pomdp,vendrov2020gradient}. Basically, there is a utility function or a score function $u(\mathbf{x}_j, \mathbf{u}_i)$ representing user $i$'s preference for item $j$. Usually, it can be written as a linear function as
\begin{equation}
  u\left(\mathbf{x}_{j} , \mathbf{u}_i\right)=\mathbf{x}_{j}^{T} \mathbf{u}_i.
\end{equation}
In a Bayesian setting, user $i$'s preference is modeled by a probabilistic distribution instead of a deterministic vector, which means that the vector $\mathbf{u}_i$ is sampled from a prior user belief $P\left(\mathcal{U}^{(i)}\right)$. Therefore, the utility of an item $j$ for a user $i$ is computed as the expectation:
\begin{equation}
  \mathbb{E}\left[u\left(\mathbf{x}_{j} , \mathbf{u}_i\right)\right]=\int_{\mathbf{u}_i \sim \mathcal{U}^{(i)}} P(\mathbf{u}_i) u\left(\mathbf{x}_{j} , \mathbf{u}_i\right) d \mathbf{u}_i.
\end{equation}
The item with the maximum expected utility for user $i$ is considered as the recommendation items:
\begin{equation}
  \arg \max _{j} \mathbb{E}\left[u\left(\mathbf{x}_{j} , \mathbf{u}_i\right)\right].
\end{equation}
Based on the utility function, the system can select some items to query. And the user belief distribution can be updated based on users' feedback. Specifically, given a user response $r_j$ to the question $q$, the posterior user belief $P(\mathbf{u}_i|q, r_j)$ can be written as:
\begin{equation}
  P(\mathbf{u}_i|q, r_j) = \frac{P\left(r_j \mid q, \mathbf{u}_i\right) P(\mathbf{u}_i)}{\int_{\mathcal{U}^{(i)}} P\left(r_j \mid q, \mathbf{u}_i\right) P(\mathbf{u}_i) d \mathbf{u}_i}.
\end{equation}
As for the query strategy, i.e., selecting which items to ask, there are different criteria. For example, \citet{boutilier2002pomdp} propose a partially observed Markov decision process (POMDP) framework as the sequential query strategy. And \citet{vendrov2020gradient} and \citet{guo2010real} use the expected value of information (EVOI) paradigm as a relatively myopic strategy to select items to query. Furthermore, the query type can be classified into two different types: 
\begin{enumerate*}[label=(\arabic*)]
\item a pairwise comparison query, in which the users are required to choose what they prefer more between two items or two item sets \cite{christakopoulou2016towards,guo2010real,sepliarskaia2018preference}; or 
\item a slate query, where users need to choose from multiple given options \citep{vendrov2020gradient}.
\end{enumerate*}

\medskip\noindent\textbf{Interactive Recommendation.}
Interactive recommendation models are mainly based on \acl{RL}. Some researchers adopt a \ac{MAB} algorithm~\cite{cikm13/ICF,christakopoulou2016towards,wang2018online}. The advantage is two-fold. First, \ac{MAB} algorithms are efficient and naturally support conversational scenarios. Second, \ac{MAB} algorithms can exploit the items that users liked before and explore items that users may like but never tried before. There are also researchers formulate the interactive recommendation as a meta learning problem which can quickly adapt to new tasks \cite{zou2020neural,lee2019melu}. A task here is to make recommendations based on several conversation histories. Meta learning methods and MAB-based methods have the capability of balancing exploration and exploitation.  We will describe it later in \mysec{EE}. 

Recently, researchers incorporate \ac{DRL} models into interactive recommender systems \cite{zhao2018recommendations,chen2019large,xian2019reinforcement,zheng2018drn,hu2018reinforcement,zou2019reinforcement,chen2019top,ie2019reinforcement,Liao18mm,pecune-2019-model,zhou2020interactive,Pseudo-Dyna-Q,wang2020text}. Unlike \ac{MAB}-based methods which usually assume the user preference is unchanged during the interaction, \ac{DRL}-based methods can model a dynamic preference and long-term utility. For example, \citet{mahmood2007learning} introduce a model-based techniques and use the policy iteration algorithm \cite{sutton2018reinforcement} to acquire an adaptive strategy. Model-free frameworks such as \ac{DQN} \cite{zhao2018recommendations,zheng2018drn,zou2019reinforcement,zhou2020interactive} and \ac{DDPG} \cite{hu2018reinforcement} are used in interactive recommendation scenarios. Most \ac{RL}-based methods often suffer from low efficiency issues and cannot handle cold-start users. \citet{zhou2020interactive} propose to integrate a knowledge graph into the interactive recommendation to solve these problems.

For more works that leverage \ac{RL} in interactive recommender systems, we refer the interested readers to the comprehensive survey conducted by \citet{afsar2021reinforcement}.

\medskip\noindent
However, directly requiring items is inefficient for building the user profile because the candidate item set is large. In real-world \ac{CRS} applications, users will get bored as the number of conversation turns increases. It is more practical to ask attribute-centric questions, i.e., to ask users whether they like an attribute (or topic/category in some works), and then make recommendations based on these attributes \cite{zhang2018towards,lei20estimation}. Therefore, the estimation and utilization of a user's preferences towards attributes become a key research issue. 

\subsection{Asking about Attributes}
\label{sec:ask_attributes}
Asking about attributes is more efficient because whether users like or dislike an attribute can significantly reduce the recommendation candidates. The challenge is to determine a sequence of attributes to ask so as to minimize the uncertainty of current user needs \cite{mirzadeh2005feature,thompson2004personalized}. The aforementioned critiquing-based methods fall into this category. Besides, there are other kinds of methods, we introduce some mainstream branches as below. 

\subsubsection{Fitting Patterns from Historical Interaction}
A conversation can be deemed as a sequence of entities including consumed items and mentioned attributes, and the objective is to learn to predict the next attribute to ask or the next item to recommend. Therefore, the sequential neural network such as the \ac{GRU} model \cite{cho2014learning} and the \ac{LSTM} model \cite{hochreiter1997long} can be naturally adopted in this setting, due to its ability to capture long and short term dependency in user behavioral patterns.

An exemplar work is the question \& recommendation (Q\&R) model proposed by \citet{christakopoulou2018q}, 
where the interaction between the system and a user is implemented as a selection system. In each turn, the system asks the user to choose one or more distinct topics (e.g., NBA, Comics, or Cooking) from the given list, and then recommends items in these topics to the user. 
It contains a trigger module to decide whether to ask a question about attributes or to make a recommendation. The triggering mechanism can be as simple as a random mechanism or can be more sophisticated, i.e., using criteria capturing the user’s state, or even be user-initiated. At the $t$-th time step, the next topic $q$ that user click can be predicted based on the user's watching history $e_{1}, \ldots, e_{T}$ as: $P\left(q \mid e_{1}, \ldots, e_{T}\right)$. After user clicking a topic $q$, the model can recommend an item $r$ based on the conditional probability written as: $P\left(r \mid e_{1}, \ldots, e_{T}, q\right)$.
Both of the two conditional probabilities are implemented as the \ac{GRU} architecture \cite{cho2014learning}. This algorithm is deployed on YouTube, for obtaining preferences from cold-start users.

\citet{zhang2018towards} propose a ``System Ask User Response'' (SAUR) paradigm. For each item, they utilize the rich review information and convert a sentence containing an aspect-value pair to a latent vector via the \ac{GRU} model. Then they adopt a memory module with attention mechanism \cite{sukhbaatar2015end,kumar2016ask,miller-etal-2016-key} to perform both the next question generation task (determining which attribute to ask) and the next item recommendation task. Again, they also develop a heuristic trigger to decide whether it is the time to display the top-$n$ recommended items to users or to keep asking questions about attributes. One limitation of the work is that the authors assume all information in reviews can support the purchasing behavior, however it is not true as users may complain certain aspects of the purchased items, e.g., a user may write ``64 Gigabytes is not enough''. Using information without discrimination will mislead the model and deteriorate the performance.

The utterances produced by the system, i.e., the questions, are constructed with predefined language patterns or templates, meaning that what the system needs to pay attention to are only the aspect and the value. This is a common setting in state-of-the-art \ac{CRS} studies because the core task here is recommendation instead of language generation \cite{christakopoulou2018q,lei20estimation,lei2020interactive}.

Note that these kinds of methods have a common disadvantage: learning from historical user behaviors cannot aid understanding the logic behind the interaction. As interactive systems, these models do not consider how to react to feedback when users reject the recommendation, i.e., they just try to fit the preferences in historical interaction and do not consider an explicit strategy to deal with different feedback.

\subsubsection{Reducing Uncertainty}

Unlike sequential neural network-based methods that do not have an explicit strategy to handle all kinds of user feedback, some studies try to build a straightforward logic to narrow down item candidates.

\medskip\noindent\textbf{Critiquing-based Methods.}
The aforementioned critiquing model is typically equipped with a heuristic tactic to elicit user preference on attributes \cite{chen2012critiquing,10.1145/3298689.3347009,luo2020deep,luo2020latent}. In traditional critiquing models, where the critique on an attribute value (e.g., ``not red'' for color or ``less expensive'' for price) is used for reconstructing the candidate set by removing the items with unsatisfied attributes \cite{chen2012critiquing,mccarthy2004thinking,smyth2004compound,viappiani2007conversational,burke1997findme,smyth2003analysis}. The neural vector-based methods take the criticism into the latent vector, which is responsible for generating both the recommended items and the explained attributes. For example, \citet{10.1145/3298689.3347009} propose an explainable neural collaborative filtering (CE-NCF) model for critiquing. They use the \ac{NCF} model \cite{he2017ncf} to encode the preference of a user $i$ for an item $j$ as a latent vector $\hat{\mathbf{z}}_{i, j}$, then $\hat{\mathbf{z}}_{i, j}$ is used for producing the rating score $\hat{r}_{i, j}$ as well as the explained attribute vector $\hat{\mathbf{s}}_{i, j}$. The attributes are composed of a set of key-phrases such as ``golden, copper, orange, black, yellow,'' and each dimension of $\hat{\mathbf{s}}_{i, j}$ corresponds to a certain attribute. When a user dislikes an attribute and critique it in real-time feedback, the system updates the explained attribute vector $\hat{\mathbf{s}}_{i, j}$ by setting the corresponding dimension to zero. Then the updated vector $\tilde{\mathbf{s}}_{i, j}$ is used to update the latent vector $\hat{\mathbf{z}}_{i, j}$ to be $\tilde{\mathbf{z}}_{i, j}$. Consequently, the recommendation score is updated to be $\tilde{r}_{i, j}$. Following this setting, \citet{luo2020deep} change the base \ac{NCF} model \cite{he2017ncf} to be a \ac{VAE} model, and this generative model can help the critiquing system have better computational efficiency, improved stability, and faster convergence.

\medskip\noindent\textbf{Reinforcement Learning-driven Methods.}
Reinforcement learning is also used in \acp{CRS} to select the appropriate attributes to ask \cite{Sun:2018:CRS:3209978.3210002,lei20estimation,lei2020interactive}. Empowered by a deep policy network, the system not only selects the attributes but also determine a controlling strategy on when to change the topic of the current conversation; we will elaborate this in \mysec{RL-in-multi-turn} where we describe how reinforcement learning helps the system form a multi-turn conversational strategy. 

\medskip\noindent\textbf{Graph-constrained Candidates.}
\begin{figure}[pos=t]
\centering
\includegraphics[width=1\linewidth]{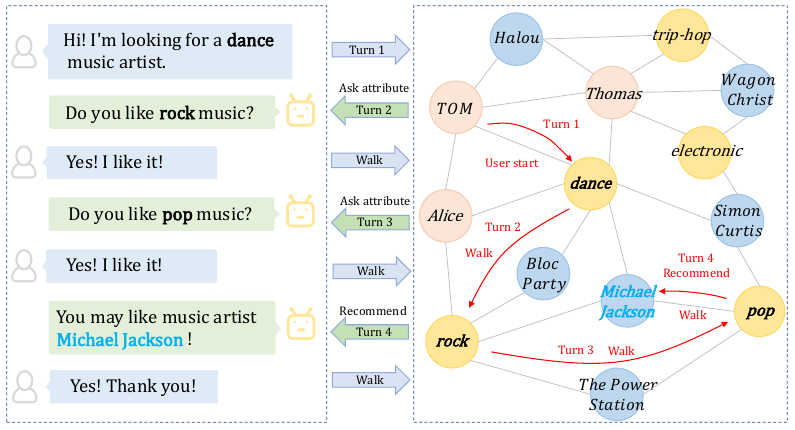}
\caption {An illustration of interactive path reasoning in the \ac{CPR} model. Credits: \citet{lei2020interactive}.}
\label{fig:CPR}
\end{figure}
Graph is a pre\-valent structure to represent relationship of different entities. It is natural to utilize graphs to sift items given a set of attributes. For example, \citet{lei2020interactive} propose an interactive path reasoning algorithm on a heterogeneous graph on which users, items, and attributes are represented as nodes and an edge connected two nodes represented a relationship between two nodes, e.g., a user purchased an item, or an item has a certain value for an attribute. With the help of the graph, a conversation can be converted to a path on the graph, as illustrated in \myfig{CPR}. The authors compare the uncertainty of preference for attributes and choose the attributes with the maximum uncertainty to ask. Here, the preference for a certain attribute is modeled by the average preference for items that have this attribute. Hence, the searching space and overhead of the algorithm can be significantly reduced by utilizing the graph information. There are other studies that apply graph neural networks (GNNs) to learn a powerful representation of both items and attributes, so the semantic information in the learned embedding vectors can help end-to-end \ac{CRS} models generate appropriate recommendations. For example, the \ac{GCN} model and its variants \cite{GCN,schlichtkrull2018modeling} are adopted on the knowledge graph in recent \ac{CRS} models \cite{chen-etal-2019-towards,zhou2020improving,xu2020user,liao2020TKDE}.

\begin{table*}[pos=t]
\tabcolsep=3pt
\scriptsize
\renewcommand\arraystretch{1.7}
\caption{Characteristics of common \ac{CRS} models in different dimensions. The strategy indicates whether the work considers an explicit strategy to control multi-turn conversations, e.g., whether to ask or recommend in the current turn.}
\label{tab:characteristics}
\begin{tabular}{llllll}
\toprule
\textbf{Asking}                      & \textbf{Asking Mechanism}                                                                         & \textbf{Basic Model}                       & \textbf{Type of User Feedback}                        & \textbf{Strategy} & \textbf{Publications}                                                                                                                                                                                                                                                              \\
\midrule
\multirow{4}{*}{\textbf{Items}}      & Exploitation \& Exploration                                                                       & Multi-armed bandit                         & Rating on the given item(s)                           & No                & \cite{cikm13/ICF,christakopoulou2016towards,zhou2020conversational,ISR2017CIKM,10.1145/3292500.3330991}                                                                                                                                                                                                                        \\ \cmidrule{2-6}
                                     & Exploitation \& Exploration                                                                       & Meta learning                              & Rating on the given item(s)                           & No                & \cite{zou2020neural,lee2019melu}                                                                                                                                                                                                                                                  \\ \cmidrule{2-6}
                                     & Maximal posterior user belief                                                                     & Bayesian methods                           & Rating on the given item(s)                           & No                & \cite{vendrov2020gradient}                                                                                                                                                                                                                                                        \\ \cmidrule{2-6}
                                     & Reducing uncertainty                                                                              & Choice-based methods                       & Choosing an item or a set of items                    & No                & \cite{loepp2014choice,jiang2014choice,graus2015improving,saavedra2016choice,iui2020navigation}                                                                                                                                                                                    \\ \midrule
\multirow{15}{*}{\textbf{Attributes}}& Exploitation \& Exploration                                                                       & Multi-armed bandit                         & Rating on the given attribute(s)                      & Yes               & \cite{zhang2020conversational,li2020seamlessly}                                                                                                                                                                                          \\ \cmidrule{2-6}
                                     & \multirow{3}{*}{Reducing uncertainty}                                                             & Bayesian approach                          & Providing preferred attribute values                  & No                & \cite{mangili2020bayesian,Wu:umap21}                                                                                                                                                                                                                                   \\ \cmidrule{3-6}
                                     & 										                                                             & Critiquing-based methods                      & Critiquing one/multiple attributes                    & No                & \begin{tabular}[c]{@{}l@{}}\cite{mccarthy2004thinking,smyth2004compound,viappiani2007conversational,burke1997findme,smyth2003analysis}\\ \cite{pu2004decision,chen2012critiquing,10.1145/3298689.3347009,luo2020deep,luo2020latent}\end{tabular} \\ \cline{3-6}
                                     & 								                                                                     & Matrix factorization              		  & Answering Yes/No for an attributes                    & No                & \cite{10.1145/3397271.3401180}                                                                                                                                                                                                                                                    \\ \cmidrule{2-6}
                                     & \multirow{2}{*}{Fitting historical patterns}                                                      & \multirow{2}{*}{Sequential neural network} & Providing preferred attribute values                  & Yes               & \cite{christakopoulou2018q,zhang2018towards}                                                                                                                                                                                                                                      \\ \cmidrule{4-6}
                                     &                                                                                                   &                                            & Providing an utterance                                & No                & \cite{nips18/DeepConv,chen-etal-2019-towards}                                                                                                                                                                                                                                     \\ \cmidrule{2-6}
                                     & \multirow{3}{*}{Maximal reward}                                                                   & \multirow{3}{*}{Reinforcement learning}    & Answering Yes/No for an attributes                    & Yes               & \cite{lei20estimation,lei2020interactive}                                                                                                                                                                                                                                         \\ \cmidrule{4-6}
                                     &                                                                                                   &                                            & \multirow{2}{*}{Providing an utterance}               & Yes               & \cite{Sun:2018:CRS:3209978.3210002,Tsumita19Dialogue,kang-etal-2019-recommendation}                                                                                                                                                                                               \\ \cmidrule{5-6}
                                     &                                                                                                   &                                            &                                                       & No                & \cite{10.1145/3394592}                                                                                                                                                                                                                                                            \\ \cmidrule{2-6}
                                     & \multirow{5}{*}{\begin{tabular}[c]{@{}l@{}}Exploring graph-constrained\\ candidates\end{tabular}} & \multirow{5}{*}{Graph reasoning}           & Answering Yes/No for an attributes                    & Yes               & \cite{lei2020interactive}                                                                                                                                                                                                                                                         \\ \cmidrule{4-6}
                                     &                                                                                                   &                                            & \multirow{2}{*}{Providing an utterance}               & Yes               & \cite{chen-etal-2019-towards,liu2020ACL}                                                                                                                                                                                                                                          \\ \cmidrule{5-6}
                                     &                                                                                                   &                                            &                                                       & No                & \cite{zhou2020improving,liao2020TKDE}                                                                                                                                                                                                                                     \\ \cmidrule{4-6}
                                     &                                                                                                   &                                            & \multirow{2}{*}{Providing preferred attribute values} & Yes               & \cite{xu2020user}                                                                                                                                                                                                                                          \\ \cmidrule{5-6}
                                     &                                                                                                   &                                            &                                                       & No                & \cite{moon-2019-opendialkg}                                                                                                                                                                                                                                     \\ \bottomrule
\end{tabular}

\end{table*}

\medskip\noindent\textbf{Other Methods.}
There are other attempts to make recommendations based on user feedback on attributes. For example, \citet{10.1145/3397271.3401180} proposed a question-driven recommender system based on an extended matrix factorization model, which merely considers the user rating data, to combine real-time feedback from users. 

The basic assumption is that if a user likes an item, then he/she will like the attributes of this item. Thereby, in each turn, the system will select the attribute that carries the maximum amount of uncertainty to ask. In other words, if an attribute is known to be shared by most items that a user likes, then it does not need to ask about this attribute. Similarly, there is no need to ask about the attributes that users dislike. Only if it is not sure whether a user likes an attribute, then asking about this attribute can provide the most amount of information. The parameters in matrices can be updated after users providing feedback. Besides, using ideas similar to aforementioned models based on asking items, \ac{MAB}-based models \cite{zhang2020conversational,li2020seamlessly} and Bayesian approaches \cite{mangili2020bayesian} are also developed in attribute-asking \acp{CRS}.

\subsection{Section Summary}
We list the common \ac{CRS} models in \mytable{characteristics}, where the models are characterized by different dimensions, which are the asking entity (item or attribute), the asking mechanism, the type of user feedback, and the multi-turn strategy that we will describe in the next section.

In most interactive recommendations \cite{Pseudo-Dyna-Q,wang2020text,zhang2019reward,ding2020hybrid} and critiquing methods \cite{chen2012critiquing,10.1145/3298689.3347009,luo2020deep,luo2020latent}, the system keeps asking questions, and each question is followed by a recommendation. This process will only terminate when users quit with either being satisfied or impatient. The setting is unnatural and will likely hurt the user experience during the interaction process. Asking too many questions may let the interaction become an interrogation. Moreover, during the early stages of interaction, when the system has not confidently modeled the user preferences yet, recommendations with low confidence should not be exposed to the user \cite{schnabel2018short}. In other words, there should be a multi-turn conversational strategy to control how to switch between asking and recommending, and this strategy should change dynamically in the interaction process.

\section{Multi-turn Conversational Strategies for CRSs}
\label{sec:multi-turn}

Question-driven methods focus on the problem of ``\emph{What to ask}'', and the multi-turn conversational strategies discussed in this section focus on ``\emph{When to ask}'' or a broader perspective, ``\emph{How to maintain the conversation}''. A good strategy cannot only make the recommendation at the proper time (with high confidence) and adapt flexibly to users' feedback, but also maintain the conversation topics and adapt to different scenarios to make users feel comfortable in the interaction.

\subsection{Conversation Strategies for Determining When to Ask and Recommend}
\label{sec:RL-in-multi-turn}
Most \ac{CRS} models do not carefully consider a strategy to determine whether to continue interrogating users by asking questions or to make a recommendation. However, a good strategy is essential in the interaction process so as to improve the user experience. The strategy can be a rule-based policy, i.e., making recommendations every $k$ turns of asking questions \cite{zhang2020conversational}, or a random policy \cite{christakopoulou2018q}, or a model-based policy \cite{christakopoulou2018q}. 

In the SAUR model \cite{zhang2018towards}, a trigger is set to activate the recommendation module when the confidence is high. The trigger is simply implemented as a sigmoid function on the score of the most probable item, i.e., if the score of the candidate item is high enough, then the recommendation step is triggered, else the system will keep asking questions.

Though straightforward and easy to control, these strategies cannot capture rich semantic information, e.g., what topics are talking about now or how deep the topics have been explored. This information can directly affect the conversation topic. Thereby, a sophisticated strategy is necessary. Recently, \acf{RL} has been adopted by many interactive recommendation models for its potential of modeling the complex environment \cite{zhao2018recommendations,chen2019large,xian2019reinforcement,zheng2018drn,zou2019reinforcement,chen2019top,ie2019reinforcement,Liao18mm,pecune-2019-model,zhang2019reward,zhou2020interactive}. Therefore, it is natural to incorporate \ac{RL} into the \ac{CRS} framework \cite{Sun:2018:CRS:3209978.3210002,lei20estimation,lei2020interactive,Tsumita19Dialogue,10.1145/3394592,kang-etal-2019-recommendation}. For instance, \citet{Sun:2018:CRS:3209978.3210002} propose a model called \ac{CRM} that uses the architecture of task-oriented dialogue system. 
In \ac{CRM}, a belief tracker is used to track the users' input, and it outputs a latent vector representing the current state of the dialogue and the user preferences that have so far been captured. Afterward, the state vector of the belief tracker is input into a deep policy network to decide whether to recommend an item or to keep asking questions. Specifically, there are $l+1$ actions: $l$ actions for choosing one facet to ask and the last one is to yield a recommendation. The deep policy network uses the policy gradient method to make decisions. Finally, the model gets rewards from the environment, which includes user feedback towards the questions and the reward from the automatic evaluation of recommendation results.

\begin{figure}[pos=t]
\centering
\includegraphics[width=1\linewidth]{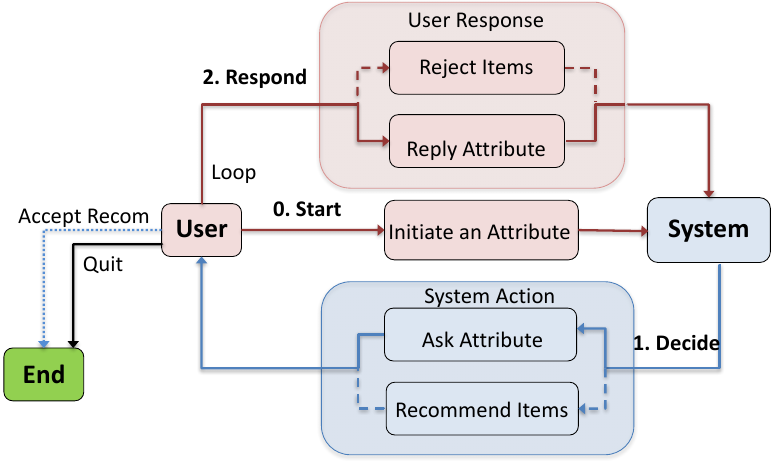}
\caption {The estimation-action-reflection workflow. Credits: \citet{lei20estimation}.}
\label{fig:ask_lei_workflow}
\end{figure}

However, the state modeled in \ac{CRM} is a latent vector capturing the information of facet-values, which is hard to interpretable. In this respect, some studies explore better ways to construct the state of \ac{RL} to make the multi-turn conversation strategy better adapt to an dynamic environment. For example, \citet{lei20estimation} propose an \ac{EAR} framework, which assumes that the model should only ask questions at the right time. The right time, in their definition, is when 
\begin{enumerate*}[label=(\arabic*)]
\item the item candidate space is small enough;
\item asking additional questions is determined to be less useful or helpful, from the perspective of either information gain or user patience; and
\item the recommendation engine is confident that the top recommendations will be accepted by the user. 
\end{enumerate*}
The workflow of the \ac{EAR} framework is illustrated in \myfig{ask_lei_workflow}, where the system has to decide whether to continue to ask questions about attributes or to make a recommendation based on available information. 
To determine when to ask a question, they construct the state of the RL model to take into account four factors:
\begin{itemize}[nosep,leftmargin=*]
    \item Entropy information of each attribute among the attributes of the current candidate items. Asking attributes with a large entropy helps to reduce the candidate space, thus benefits finding desired items in fewer turns.
    
    \item User preference on each attribute. 
    The attribute with a high predicted preference is likely to receive positive feedback, which also helps to reduce the candidate space.
    
    \item Historical user feedback. If the system has asked about a number of attributes for which the user gives approval, it may be a good time to recommend.
    
    \item Number of rest candidates. If the candidate list is short enough, the system should turn to recommend to avoid wasting more turns.
\end{itemize}
\noindent%
Building on these vectors capturing the current state, the RL model learns the proper timing to ask or recommend, which is more intelligent than a fixed heuristic strategy. 

During the conversation, the recommendation module takes the items in the previous list of recommendations that are not chosen by users as the negative samples. However, \citet{lei20estimation} mention that this setting deteriorates the performance of the recommendation results. The reason, as they analyze it, is that rejecting the produced attribute does not mean that the user dislikes it: maybe the user does like it but overlooks it or just wants to try other new things.

Furthermore, \citet{lei2020interactive} extend the \ac{EAR} model by proposing the \ac{CPR} model. By integrating the knowledge graph consisted of users,  items, and attributes, they model conversational recommendation as an interactive path reasoning problem on the graph. A toy example of the generated conversation of the \ac{CPR} model is shown in \myfig{CPR}. Unlike the \ac{EAR} model where the attributes to be asked are selected irregular and unpredictable from all attribute candidates, \ac{CPR} chooses attributes to be asked and items to be recommended strictly following the paths on the knowledge graph, which renders interpretable results.

In terms of the timing to ask or recommend, CRP makes an important improvement: the action space of the RL policy is only two --- asking an attribute or making item recommendations. 
This largely reduces the difficulty of learning the RL policy. 
The \ac{CPR} model is much more efficient than the \ac{EAR} model due to the fact that the searching space of attributes in \ac{CPR} is constrained by the graph. The integration of knowledge improves the multi-turn conversational reasoning ability.

\subsection{Conversation Strategies from A Broader Perspective}

Although learning from the query-answering interactions can enable the system to understand and respond to human query directly, the system still lacks intelligence. One reason is that most \ac{CRS} models assume that users always bear in mind what they want, and the task is to obtain the preference through asking questions. However, users who resort to recommendation might not have a clear idea about what they really want. Just like a human asks a friend for suggestions on restaurants. Before that, he may not have a certain target in mind, and his decision can be affected by his friend's opinions. 
Therefore, \acp{CRS} should not only ask clarification questions and interrogate users, but also take responsibility for leading the topics and affecting users' mind. 
Towards this objective, some studies try to enrich \acp{CRS} certain personalities or endow \acp{CRS} the ability to lead the conversation, which can make the dialogues more attractive and more engaging. These efforts can also be found in the field of proactive conversation \cite{AAAI1816104,wu-etal-2019-proactive,balaraman2020proactive}.

\subsubsection{Multi-topic Learning in Conversations}

Borrowing the idea from the proactive conversation, \citet{liu2020ACL} present a new task which places conversational recommendation in the context of multi-type dialogues. In their model, the system can proactively and naturally lead a conversation from a non-recommendation dialog (e.g., question answering or chitchat) to a recommendation dialog, taking into account the user’s interests and feedback. 
And during the interaction, the system can learn to flexibly switch between multiple goals.
To address this task, they propose a \ac{MGCG} framework, which consists of a goal planning module and a goal-guided responding module. The goal-planning module can conduct dialog management to control the dialog flow, which takes recommendation as the main goal and complete the natural topic transitions as the short-term goals. Specifically, given a user's historical utterances as context $X$ and the last goal $g_{t-1}$, the module estimates the probability of changing the goal $g_{t}$ of the current task as $P_{GC}(g_{t}\neq g_{t-1}|X,g_{t-1})$. The goal $g_{t}$ of the current task is changed when the probability $P_{GC}>0.5$ and remains to be $g_{t-1}$ if $P_{GC}\le 0.5$. Based on the current goal, the framework can produce responses from an end-to-end neural network.

Learning a multi-type conversational model requires a dataset that supports multi-type dialogues. Therefore, \citet{liu2020ACL} create a dataset, denoted as DuRecDial, with various types of interaction. In DuRecDial, two human workers are asked to conduct the conversation based on a given profile, which contains the information of age, gender, occupation, preferred domains, and entities. The workers must produce utterances that are consistent with their given profiles, and they are encouraged to produce utterances with diverse goals, e.g., question answering, chitchat, or recommendation. Then these dialogue data are labeled with goals and goal descriptions by templates and human annotation. 

Further, \citet{zhou2020topicguided} release a topic-guided conversational recommendation dataset. They collect the review data from Douban Movie \footnote{\url{https://movie.douban.com/}}, a movie review website, to construct the recommended movies, topic threads, user profiles, and utterances. And they associate each movie with the concepts in ConceptNet \cite{ConceptNet}, a commonsense knowledge graph, for providing rich topic candidates. Then they use rules to generate multi-turn conversations with diverse topics based on the user profile and topic candidates. Based on the proposed dataset, a new task of topic-guided conversational recommendation is defined as follows: given the user profile $P_u$, user interaction sequence $I_u$, historical utterances ${s_1, \ldots , s_{k-1}}$, and corresponding topic sequence $\left\{t_{1}, \ldots, t_{k-1}\right\}$, the system should: 
\begin{enumerate*}[label=(\arabic*)]
\item predict the next topic $t_k$, or
\item recommend the movie $i_k$, and finally 
\item produce a proper response $s_k$ about the topic and with persuasive reasons.
\end{enumerate*}

\begin{table*}[pos=t]
\tabcolsep=5pt
\footnotesize
\renewcommand\arraystretch{1.3}
\caption{The commonly used multi-turn strategies in \acp{CRS}.}
\label{tab:multi-turn-strategy}
\begin{tabular}{@{}lcccl@{}}
\toprule
\textbf{Main Mechanism}                    & \textbf{Asking Method}    & \textbf{When to ask and recommend}                       & \textbf{Determining $X$ and $Y$} & \textbf{Publications} \\ \midrule
\multirow{5}{*}{\textbf{Asking questions}} & \multirow{4}{*}{Explicit} & Asking $1$ turn; recommending $1$ turn                 & Fixed                            & \cite{christakopoulou2018q,10.1145/3292500.3330991}     \\ \cmidrule(l){3-5} 
                                           &                           & \multirow{2}{*}{Asking $X$ turn(s); recommending $1$ turn} & Fixed                            & \cite{10.1145/3397271.3401180}                 \\ \cmidrule(l){4-5} 
                                           &                           &                                               & Adaptive                         & \cite{Sun:2018:CRS:3209978.3210002}                     \\ \cmidrule(l){3-5} 
                                           &                           & Asking $X$ turn(s); recommending $Y$ turn(s)                  & Adaptive                         & \cite{zhang2018towards,lei20estimation,lei2020interactive,li2020seamlessly,xu2021adapting} \\ \cmidrule(l){2-5} 
                                           & Implicit                  & Contained in natural language                                 & Adaptive                         & \cite{nips18/DeepConv,chen-etal-2019-towards,zhou2020improving,zhou2020topicguided}                      \\ \midrule
\multicolumn{3}{l}{\textbf{Leading diverse topics or explore special abilities}}                                                                     &                                  & \cite{liu2020ACL,zhou2020topicguided,rosset2020leading,lewis-etal-2017-deal,wang-etal-2019-persuasion}                      \\ \bottomrule
\end{tabular}
\end{table*}

\subsubsection{Special Ability: Suggesting, Negotiating, and Persuading}
There are miscellaneous tasks beyond the preference elicitation and recommendation for an intelligent interactive system, which require the \ac{CRS} to possess different abilities to react in different scenarios. This is a high-level and abstract requirement. A lot of effort have put into helping the machine improve the topic's guiding ability. For instance, in conversational search \cite{voskarides-2020-query,terhoeve-2020-conversations,ren-2020-conversations-arxiv,vakulenko-2020-conversational-arxiv,vakulenko-2021-large-scale-arxiv,ren-2021-wizard}, where traditional work has mainly attempted to better understand a user’s information needs by resolving ambiguity, the conversational search engine aims to lead the conversation with questions that a user may want to ask in the next step. For example, if a user queried ``Nissan GTR Price,'' then the system can provide question suggestions include those that help the user complete a task (``How much does it cost to lease a Nissan GT-R?''), weigh options (``What are the pros and cons of the Nissan GT-R?''), explore an interesting related topic (``Is the Nissan GT-R the ultimate streetcar?''), or learn more details (``How much does 2020 Nissan GTR cost?'') \cite{rosset2020leading}. These question suggestions can lead the user to an immersive search experience with diverse and fruitful future outcomes.

In addition, \citet{lewis-etal-2017-deal} propose a system that is capable of engaging in the negotiations with users. They define the problem as an allocation problem: there are some items that need to be allocated to two people, where each item has a different value to a different person and people do not know the value of others. Hence, the two people have to converse and negotiate with each other to reach an agreement about the division of these items. Instead of optimizing relevance-based likelihood, the model should pursue a maximal profit for both parties. The authors use \ac{RL} to tackle this problem. And they interleave \ac{RL} updates with supervised updates to avoid that the models diverges from human language. 

\citet{wang-etal-2019-persuasion} develop a model that tries to persuade users to take certain actions, which is very promising for conversational recommendation. They train the model, according to conversational contexts, to learn and predict the $10$ persuasion strategies (e.g., logical appeal or emotion appeal) used in the corpus. And they analyze which strategies are better conditioned on the background (personality, morality, value systems, willingness) of the user being persuaded. 

Though some of these efforts are applied to specific application scenarios in dialogue systems, these techniques can be adopted in the multi-turn strategy in \acp{CRS} and thus push the development of \acp{CRS}.

\subsection{Section Summary}
The multi-turn conversation strategies of \acp{CRS} discussed in this section are summarized in \mytable{multi-turn-strategy}. The main focus of the conversation strategy is to determine when to elicit user preference by asking questions and when to make recommendations. As a recommendation should only be made when the system is confident, an adaptive strategy can be more promising compared to a static one. Besides this core function, we introduce some strategies from a broader perspective. These strategies can extend the capability of \acp{CRS} by means of leading multi-topic conversations \cite{liu2020ACL,zhou2020topicguided} or showing special ability such as suggesting \cite{rosset2020leading}, negotiating \cite{lewis-etal-2017-deal}, and persuading \cite{wang-etal-2019-persuasion}.
\section{Dialogue Understanding and Generation in CRSs}
\label{sec:NLUNLG}

An important direction of \acp{CRS} is to converse with humans in natural languages, thus understanding human intentions and generating human-understandable responses are critical. However, most \acp{CRS} only extract key information from processed structural data and present the result via rule-based template responses \cite{zhang2018towards,10.1145/3397271.3401180,lei20estimation,lei2020interactive}. This not only requires lots of labor to construct the rule or template but also make the result rely on the preprocessing. It also hurt user experience as the constrained interaction is unnatural in real-world applications. Recently, we have witnessed the development of end-to-end learning frameworks in dialogue systems, which have been studying for years to automatically handle the semantic information in raw natural language \cite{gao2019neural, lei2018sequicity,jin2018explicit}. We will introduce these \ac{NLP} technologies in dialogue systems and describe how they help \acp{CRS} understand user intention and sentiment and generate meaningful responses.

\subsection{Dialogue Understanding}
Understanding users' intention is the key requirement for the user interface of a \ac{CRS}, as downstream tasks, e.g., recommendation, rely heavily on this information. However, most \acp{CRS} pay attention to the core recommendation logic and the multi-turn strategy, while they circumvent extracting user intention from raw utterances and requires the preprocessed input such as rating scores \cite{cikm13/ICF,christakopoulou2016towards,zou2020neural,lee2019melu}, YES/NO answers \cite{10.1145/3397271.3401180,lei20estimation,lei2020interactive}, or another type of value or orientation \cite{christakopoulou2018q,zhang2018towards} towards the queried items or attributes. This is unnatural in real-life human conversation and imposes constraints on user expression. Thereby, it is necessary to develop methods to extract semantic information in users' raw language input, either in an explicit or implicit way. 

We introduce how dialogue systems use \ac{NLP} technologies to address this problem and give the examples of \acp{CRS} that use these technology to understand user intention.

\subsubsection{Slot Filling}

A common way used in dialogue systems to extract useful information is to predefine some aspects of interest and use a model to fill out the values of these aspects from users' input, a.k.a, slot filling \cite{deng2012use,deoras2013deep,yao2013recurrent,mesnil2013investigation,yao2014spoken,pecune2020socially}. 
\citet{Sun:2018:CRS:3209978.3210002} first consider extracting the semantic information from the raw dialogue in \acp{CRS}. They propose a belief tracker to capture the facet-value pairs, e.g., (\emph{color, red}), from user utterances. Specifically, given a user utterance $e_t$ at time step $t$, the input to the belief tracker is the n-gram vector $\mathbf{z}_{t}$, which is written as $\mathbf{z}_{t} = \text{n-gram}(e_t)$, where the dimension of $\mathbf{z}_{t}$ is the corpus size. This means that only the positions corresponding to the words in utterance $e_t$ are set to $1$, other positions will be set to $0$. Suppose there are $K$ types of facet-value pairs, for a given facet $m \in \{1,2,\ldots,K\}$, the user's sequential utterances $\mathbf{z}_{1}, \mathbf{z}_{2}, \cdots, \mathbf{z}_{t}$ are encoded by a \ac{LSTM} model \cite{hochreiter1997long} to learn the latent vector $f_m$ for this facet $m$. The size of vector $f_m$ is set to the number of values, e.g., the number of available colors. The vector $f_m$ capturing the facet-value information will be used in the recommendation module and policy network later. Besides, \citet{10.1145/3394592,Tsumita19Dialogue} also employ \ac{RNN}-based methods to extract the facet-value information as input for in downstream tasks in their \acp{CRS}.

However, explicitly modeling semantic information as aspect-value pairs can be a limitation in some scenarios where it is difficult and also unnecessary to do that. Besides, aspect-value pairs cannot precisely express information such as user intent or sentiment. Therefore, some recent \acp{CRS} use end-to-end neural frameworks to implicit learning the representation of users' intentions and sentiment.

\subsubsection{Intentions and Sentiment Learning}

Neural networks are famous for extracting features automatically, so it can be used to extract users' intentions and sentiment in \acp{CRS}. An classic example in \acp{CRS} is the end-to-end framework that proposed by \citet{nips18/DeepConv}, which takes the user's raw utterances as input and directly produces the responses in the interaction. They collect the REDIAL dataset \footnote{\url{https://redialdata.github.io/website/}} through the crowdsourcing platform Amazon Mechanical Turk (AMT) \footnote{\url{https://www.mturk.com/}}. They pair up AMT workers and give each of them a role. The movie seeker has to explain what kind of movie he/she likes, and asks for movie suggestions. The recommender tries to understand the seeker’s movie tastes and recommends movies. All exchanges of information and recommendations are made using natural language; every movie mention is tagged using the ``$@$'' symbol to let the machine know it is a named entity. In this way, the dialogues in the REDIAL data contain the required semantic information that can help the model learn to answer users with recommendations and reasonable explanations. 
In addition, three questions are asked to provide labels for supervised learning: 
\begin{enumerate*}[label=(\arabic*)]
\item Whether the movie was mentioned by the seeker, or was a suggestion from the recommender (``suggested'' label).
\item Whether the seeker has seen the movie (``seen'' label): one of \textit{Have seen it}, \textit{Haven’t seen it}, or \textit{Didn’t say}.
\item Whether the seeker liked the movie or the suggestion (``liked'' label): one of \textit{Liked}, \textit{Didn’t like}, \textit{Didn’t say}.
\end{enumerate*}
The three labels are collected from both the seeker and the recommender.

In this way, although the facet-value constraints are removed, all kinds of information including mentioned items and attributes, user attitude, and user interest are preserved and labeled in the raw utterance. And the \ac{CRS} model needs to directly learn users' sentiment (or preferences), and it will make recommendations and generate responses based on the learned sentiment. The deep neural network-based model consists of four parts: 
\begin{enumerate*}[label=(\arabic*)]
\item A hierarchical recurrent encoder implemented as a bidirectional \ac{GRU} \cite{cho2014learning} that transforms the raw utterances into a latent vector with the key semantic information remained. 
\item At each time a movie entity is detected (with the ``$@$'' identifier convention), an \ac{RNN} model is instantiated to classify the seeker’s sentiment or opinion regarding that entity.
\item An autoencoder-based recommendation module that takes the sentiment prediction as input and produces an item recommendation. 
\item A switching decoder generating the response and deciding whether the name of the recommended item is included in the response. 
\end{enumerate*}
The model generates a complete sentence that might contain a recommended item to answer each user's utterance.

Beside using the \ac{RNN}-based neural networks, there are some \acp{CRS} that adopt the convolutional neural network (CNN) model \cite{10.1145/3394592,liu2020ACL}, which has been proven to be very effective for modeling the semantics from raw natural language \cite{kim-2014-convolutional}. 
However, deep neural networks are often criticized to be non-transparent and hard to interpret \cite{buhrmester2019analysis}. It is not clear how the deep language models can help \acp{CRS} in understanding user needs.

In order to answer this question, \citet{bertCRS2020recysc} investigate the \ac{BERT} \cite{bert2019acl}, a powerful technology for \ac{NLP} pre-training developed by Goo\-gle, to analyze whether its parameters can capture and store semantic information about items such as books, movies, and music for \acp{CRS}. The semantic information includes two kinds of knowledge needed for conducting conversational search and recommendation, namely content-based and col\-laborative-based knowledge. Content-based knowledge is knowledge that requires the model to match the titles of items with their content information, such as textual descriptions and genres. In contrast, col\-laborative-based knowledge requires the model to match items with similar ones, according to community interactions such as ratings. 
The authors use the three probes on the \ac{BERT} model (i.e., tasks to examine a trained model regarding certain properties) to achieve the goal. 
And the result shows that both col\-laborative-based and content-based knowledge can be learned and remembered. Therefore, the end-to-end language model has potential as part of \ac{CRS} models to interact with humans directly in real-world applications with complex contexts.

\subsection{Response Generation}
A natural language-based response of a \ac{CRS} should at least meet two levels of standards. The lower level standard requires the generated language to be proper and correct; the higher level standard requires the response contains meaningful and useful information about recommended results.

\subsubsection{Generating Proper Utterances in Natural Language}

Many \acp{CRS} use template-based methods to generate responses in conversations \cite{Sun:2018:CRS:3209978.3210002,lei20estimation,lei2020interactive}. However, template-based methods suffer from producing repetitive and inflexible output, and it require intense manual work. Besides, template-based responses could make users uncomfortable and hurt user experience. Hence, it is important to automate the response generation in \acp{CRS} to produce proper and fluent responses. This is also the objective of dialogue systems, so we introduce two veins of technologies for producing responses in dialogue systems:

\medskip\noindent\textbf{Retrieval-based Methods.}
The basic idea is to retrieve the appropriate response from a large collection of response candidate. This problem can be formulated as a matching problem between an input user query and the candidate responses. The most straightforward method is to measure the inner-product of the feature vectors representing a query and a response \cite{wu2018deep}. A key challenge is to learn a proper feature representation \cite{wu2018deep}. One strategy is to use neural networks to learn the representation vectors from user query and candidate response, respectively. Then, a matching function is used to combine the two representations and output a matching probability \cite{hu2014convolutional,tan2016improved,qiu2015convolutional,feng2015applying,wang2016inner}. An alternative strategy, in contrast, is to combine the representation vectors of query and response first, and then a neural method is used on the combined representation pair to further learn the interaction \cite{wang2016learning,10.5555/3016100.3016298,10.5555/3016100.3016292,lu2013deep}. These two strategies have their own advantages: the former is more efficient and suitable for online serving, while the latter is better at efficacy since the matching information is sufficiently preserved and mined \cite{wu2018deep}. 

\medskip\noindent\textbf{Generation-based Methods.}
Unlike retrieval-based methods, which select existing responses from a database of template response, generation-based methods directly produce a complete sentence from the model. The basic generation model is a recurrent sequence-to-sequence model, which sequentially feeds in each word in the query as input, and then generates the output word one by one \cite{sutskever2014sequence}.
Compared to retrieval-based methods, generation-based methods have some challenges. First, the generated answer is not guaranteed to be a well-formed natural language utterance \cite{yan2016learning}. Second, even though the generated response may be grammatically correct, we can still distinguish a machine-generated utterance from a human-generated utterance, since the machine response lacks basic commonsense \cite{young2018augmenting,zhou2018commonsense,ren-2020-thinking}, personality \cite{ijcai2018-595,zheng2020pre}, emotion \cite{zhou2018emotional}, and the ability to perceive user profiles \cite{pei-2021-cooperative}. Even worse, generation models are prone to produce a safe answer, such as ``OK,'' ``I don't understand what you are talking about,'' which can fit in almost all conversational contexts but would only hurt the user experience \cite{li-etal-2016-diversity,qiu-etal-2019-training}. \citet{ke-etal-2018-generating} propose to explicitly control the function of the generated sentence, for example, for the same user query, the system can answer with different tones: The interrogative tone can be used to acquire further information; the imperative tone is used to make requests, directions, instructions or invitations to elicit further interactions; and the declarative tone is commonly used to make statements or explanations. Another problem is how to evaluate the generated response, since there is no standard answer; we will further discuss this in \mysec{evaluation}.

Researchers borrow the ideas from dialog systems and apply the technologies in the user inferface of \acp{CRS}. For instance, \citet{nips18/DeepConv} generate responses by a decoder where a \ac{GRU} model \cite{cho2014learning} decodes the context from the previous component (i.e., predicted sentiment towards items) to predict the next utterance step by step. \citet{liu2020ACL} adopt the responding model in the work of \citet{wu-etal-2019-proactive} and propose both a retrieval-based model and a generation-based model to produce responses in their \ac{CRS}. 

However, a correct sentence does not mean it can fulfill the task of recommendation; at least the name of the recommended entity should be mentioned in generated sentences.
Hence, \citet{nips18/DeepConv} use a switch to decide whether the next predicted word is a movie name or an ordinal word; \citet{liu2020ACL} introduce an external memory module for storing all related knowledge, making the models select appropriate knowledge to enable proactive conversations. Besides, there are other efforts to guarantee the generated responses should not only be proper and accurate but also be meaningful and useful.

\subsubsection{Incorporating Recommendation-oriented Information}

There is a major limitation \acp{CRS} that use the end-to-end frameworks as the user interface: only items mentioned in the training corpus have a chance of being recommended since items that have never been mentioned are not modeled by the end-to-end model. Therefore, the performance of this method is greatly limited by the quality of human recommendations in the training data. To overcome this problem, \citet{chen-etal-2019-towards} propose to incorporate domain knowledge to assist the recommendation engine. 
The incorporation of a knowledge graph mutually benefits the dialogue interface and the recommendation engine in the \ac{CRS}.
\begin{enumerate*}[label=(\arabic*)]
\item the dialogue interface can help the recommender engine by linking related entities in the knowledge graph; the recommendation model is based on the R-GCN model \cite{schlichtkrull2018modeling} to extract information from the knowledge graph; 
\item the recommender system can also help the dialogue interface: by mining words with high probability, the dialogue can connect movies with some biased vocabularies, thus it can produce consistent and interpretable responses.
\end{enumerate*}

Following this line, \citet{zhou2020improving} point out the remaining problems in the dialogue interface in \acp{CRS}. Although \citet{chen-etal-2019-towards} have introduced an item-oriented knowledge graph to enable the system to understand the movie-related concepts, the system still cannot comprehend some words in the raw utterances. For example, ``thriller'', ``scary'', ``good plot''.  In essence, the problem originates from the fact that the dialog component and the recommender component correspond to two different semantic spaces, namely word-level and entity-level semantic spaces. Therefore, \citet{zhou2020improving} incorporate and fuse two special knowledge graphs, i.e., a word-oriented graph (ConceptNet \cite{ConceptNet}), and an item-oriented graph (DBpedia \cite{bizer2009dbpedia}), to enhance understanding semantics in both the components. 
The representations of the same concepts on the two knowledge graphs are forced to be aligned with each other via the \ac{MIM} technique \cite{velickovic2019deep,yeh-chen-2019-qainfomax}. Furthermore, a self-attention-based recommendation model is proposed to learn the user preference and adjust the representation of corresponding entities on the knowledge graph. Then, equipped with these representations containing both semantics and users' historical preferences, the authors use an encoder-decoder model to extract user intention from the raw utterances and directly generate the responses containing recommended items.

Besides, some researchers try to improve the diversity or explainability of generated responses in \acp{CRS}. For example, \citet{liu2020ACL} propose the multi-topic learning that can handle diverse dialogue types in \acp{CRS}. To enhance the interpretability of \acp{CRS}, \citet{chen2020ijcai} design an incremental multi-task learning framework to integrate review comments as side information. Hence, the \ac{CRS} can simultaneously produce a recommendation as well as a sentence as an explanation, e.g., ``I recommend \textit{Mission Impossible}, because it is by far the best of the action series.'' Moreover, \citet{luo2020deep} use a \ac{VAE}-based architecture to learn a latent representation for generating recommendations and fitting user critiquing. Therefore, their model can better understand users' intentions from users' raw comments, and thus can generate more interpretable responses. \citet{gao2020meaningful} consider attributes and review information and rewrite a coherent and meaningful answer from a selected prototype answer, which can address the safe answer problem in the response \cite{li-etal-2016-diversity,qiu-etal-2019-training}.


\begin{table}[]
\caption{Mechanisms of language understanding and generation in \acp{CRS}.}
\label{tab:language}
\tabcolsep=7pt
\renewcommand\arraystretch{1.3}
\begin{tabular}{@{}cc@{}}
\toprule
\textbf{Forms of Input \& Output}                                                          & \textbf{Publications}                                                                                                                                                                                                                                          \\ \midrule
\begin{tabular}[c]{@{}c@{}}Pre-annotated Input \&\\ Template-based Output\end{tabular} & \begin{tabular}[c]{@{}c@{}}\cite{cikm13/ICF,10.1145/3397271.3401180,loepp2014choice,zhang2018towards,Sun:2018:CRS:3209978.3210002},\\ \cite{christakopoulou2016towards,christakopoulou2018q,lei20estimation,lei2020interactive,li2020seamlessly,HOOPS}\end{tabular} \\ \midrule
\begin{tabular}[c]{@{}c@{}}Raw Language Input \& \\ Natural Language Generation\end{tabular}        & \begin{tabular}[c]{@{}c@{}}\cite{10.1145/3394592,nips18/DeepConv,chen-etal-2019-towards},\\ \cite{zhou2020improving,ma2020bridging,liu2020ACL}\end{tabular}                                                                                                   \\ \bottomrule
\end{tabular}
\end{table}

\subsection{Section Summary}
In \mytable{language}, we classify \acp{CRS} into two classes in terms of the forms of input and output. Generally, interactive recommendations \cite{Pseudo-Dyna-Q,wang2020text,zhang2019reward,ding2020hybrid}, critiquing methods \cite{chen2012critiquing,10.1145/3298689.3347009,luo2020deep,luo2020latent}, and \acp{CRS} focusing on the multi-turn conversation strategy \cite{christakopoulou2016towards,christakopoulou2018q,lei20estimation,lei2020interactive,li2020seamlessly} are prone to use the pre-annotated input and rule-based or template-based output; dialogue systems \cite{young2018augmenting,zhou2018commonsense,gao2020meaningful} and \acp{CRS} caring about the dialogue ability \cite{nips18/DeepConv,chen-etal-2019-towards,zhou2020improving} are more likely to use raw natural language as input and automatically generate responses. In the future, user understanding and response generation in \acp{CRS} will remain a critical research field, as they serve as the interface of \acp{CRS} and directly impact the user experience. 

\section{Exploration-Exploitation Trade-offs}
\label{sec:EE}

One challenge of \acp{CRS} is to handle the cold-start users that have few historical interactions. A natural way to tackle this is through the idea of the Exploration-Exploitation (E\&E) trade-off. With exploitation, the system takes advantage of the best option that is known; with exploration, the system takes some risks to collect information about unknown options. In order to achieve long-term optimization, one might make a short-term sacrifice. In the early stages of E\&E, an exploration trial could be a failure, but it warns the model to not take that action too often in the future. Although the E\&E trade-off is mainly used for the cold-start scenario in \acp{CRS}, it can also be used for improving the recommendation performance for any users (including cold users and warm-up users) in recommendation systems.

\ac{MAB} is a classic problem formulated to illustrate the E\&E trade-off, and many algorithms have been proposed to solve the problem. In \acp{CRS}, the \ac{MAB}-based algorithms are introduced to help the system improve its recommendation. Besides, there are also \acp{CRS} that use meta-learning to balance E\&E. We first introduce \ac{MAB} and common \ac{MAB}-based algorithms in recommender systems, then we present examples how \acp{CRS} balance E\&E in their models.

\subsection{Multi-Armed Bandits In Recommendation}
We first introduce the general \ac{MAB} problem and the classic methods to solve it, then we introduce how recommender systems use \ac{MAB}-based methods to achieve the E\&E balance.

\subsubsection{Introduction to Multi-Armed Bandits}

\ac{MAB} is a classic problem that well demonstrates the E\&E di\-lemma \cite{Katehakis1987TheMB,auer2002finite}. The name comes from the story where a gambler at a row of slot machines (each of which is known as a ``one-arm bandit'') wants to maximize his expected gain and has to decide which machines to play, how many times to play each machine, in which order to play them, and whether to continue with the current machine or try a different machine. The problem is difficult because all of the slot machines are black boxes, whose properties, i.e., the probability of winning, can only be estimated by the rewards observed in previous experiments.

\begin{figure}[pos=b]
\centering
\includegraphics[width=1\linewidth]{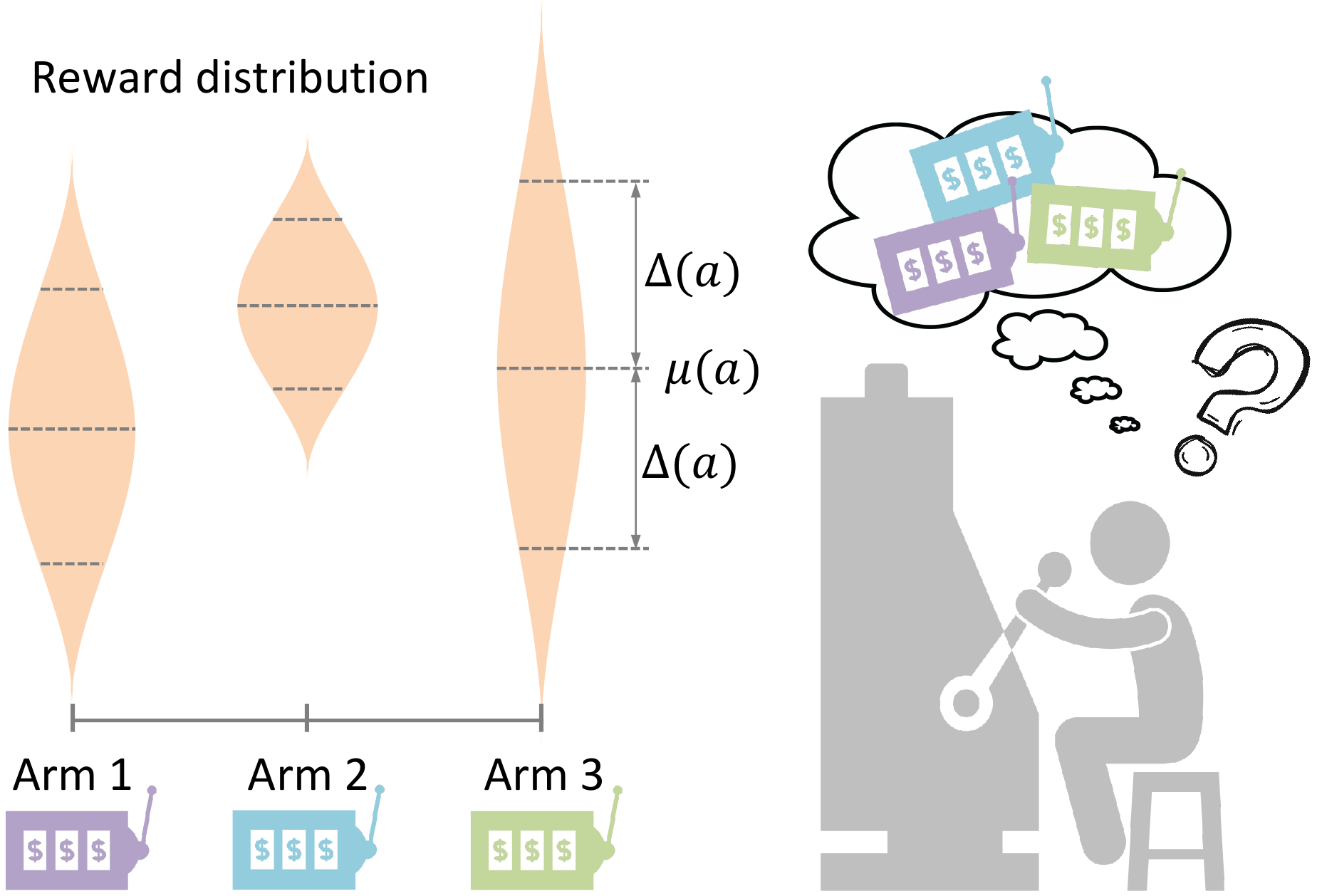}
\caption {An illustration of the \acl{MAB} problem.}
\label{fig:bandit_example}
\end{figure}

Formally, the problem is to maximize the cumulative reward $\sum_{t=1}^T r_{a, t}$ after $T$ rounds of arm selection. Here, $r_{a, t}$ is the reward with arm $0 \leq a \leq K$ selected at trial $t$, $K$ is the total number of arms. \myfig{bandit_example} illustrates an example in which a gambler decides which arm to choose now. For a certain arm, a reward distribution is estimated based on previous experiment results. The gambler can, naturally, select to exploit the second arm which has the maximal mean reward $\mu(a)$. Or, he can take some risks to explore the other arm, e.g., the third arm, which has a higher uncertainty $\Delta(a)$ and thus has the maximal upper confidence bound (UCB) of the reward $\mu(a) + \Delta(a)$. After each time he plays an arm, the new reward value is observed, and the estimated reward distribution of this arm can be updated accordingly. With exploration, the gambler hopes to find the potential arms that have higher rewards, though it can also end up in lower rewards. In any case the gambler has a better estimation of the rewards of those arms.

Equivalently, the problem can also be formulated as minimizing the regret function, which is the difference between the theoretically optimal expected cumulative reward and the estimated expected cumulative reward:
\begin{equation}
\mathbf{E}\left[\sum_{t=1}^{T} r_{t, a^{*}}\right]-\mathbf{E}\left[\sum_{t=1}^{T} r_{t, a}\right],
\end{equation}
where $a^{*}$ is the theoretically optimal arm with the maximum expected reward at all times.

The commonly used bandit strategies include the greedy strategy, i.e., the exploit-only strategy that always selects the arm with the current estimated highest reward; the random strategy, i.e., a trivial explore-only strategy; and $\epsilon$-greedy, which mixes the greedy and random strategies via a trigger with probability $\epsilon$. Other classic models include Upper Confidence Bound (UCB) \cite{auer2002using,auer2002finite} and Thompson Sampling (TS) \cite{chapelle2011empirical}  which are introduced next.

\subsubsection{Recommendation via MAB-based Methods}

As the classic algorithm for E\&E trade-offs, \ac{MAB}-based models can be seamlessly plugged into the online recommendation setting \cite{zeng2016online,zheng2018drn}, interactive recommendation \cite{cikm13/ICF,wang2017factorization}, and \acp{CRS} \cite{christakopoulou2016towards,zhang2020conversational,li2020seamlessly}. In the online or interactive recommendation tasks, the system aims to recommend the optimal item(s) according to users' previous feedback. This process can be deemed as a \ac{MAB} problem, where each arm corresponds to an item. Therefore, the classical \ac{MAB}-based methods can be plugged in this situation.

However, traditional bandit methods only consider treating items as independent arms and ignore the item features \cite{li2010contextual}. Directly estimating each item's probability of being chosen based on the accumulated rewards is rather inefficient due to a large number of items. In recommendation, there is a rich set of features on users and items, and whether a user $u_t$ would choose item $a_t$ can be predicted by the features of both $u_t$ and $a_t$. Motivated by this, \citet{li2010contextual} propose a linear contextual bandit model called LinUCB, which is the first bandit model that considers the contextual information (i.e., user/item features) in recommendation systems. 

For each trial $t$, they assume the expected reward $r_t$ of a user $u_t$ choosing an arm (item) $a_t$ is linear in its $d$-dimensional feature vector $\mathbf{x}_{u_t,a_t}$ with the unknown coefficient vector $\mathbf{\theta}^*_a$ (which is determined on this arm $a_t$ rather than other arms); namely, for all trial $t$,
\begin{equation}
\mathbf{E}\left[r_{t, a} \mid \mathbf{x}_{u_t,a_t}\right]=\mathbf{x}_{u_t,a_t}^{\top} \boldsymbol{\theta}_{a}^{*},
\end{equation}
where the feature vector $\mathbf{x}_{u_t,a_t}$ summarizes information of both user $u_t$ and arm (item) $a_t$, and is referred to as the context. The coefficients $\boldsymbol{\theta}_{a}^{*}$ can be learned from the historical interactions and feedback. 
Specifically, let $\mathbf{D}_a$ be a design matrix of dimension $m \times d$ at trial $t$, e.g., $m$ contexts that are observed previously for arm $a$, and $\mathbf{c}_t \in \mathbb{R}^{m}$ be the corresponding reward vector, the coefficients  $\boldsymbol{\theta}_{a}^{*}$ are estimated by applying ridge regression to the training data $\left(\mathbf{D}_{a}, \mathbf{c}_{a}\right)$ as:
\begin{equation*}
\hat{\boldsymbol{\theta}}_{a}=\left(\mathbf{D}_{a}^{\top} \mathbf{D}_{a}+\mathbf{I}_{d}\right)^{-1} \mathbf{D}_{a}^{\top} \mathbf{c}_{a},
\end{equation*}
where $\mathbf{I}_{d}$ is the $d \times d$ identity matrix.  When components in $\mathbf{c}_{a}$ are independent conditioned on corresponding rows in $\mathbf{D}_{a}$, it can be shown that with probability at least $1-\delta$,
\begin{equation*}
\left|\mathbf{x}_{u_t,a_t}^{\top} \hat{\boldsymbol{\theta}}_{a}-\mathbf{E}\left[r_{t, a} \mid \mathbf{x}_{u_t,a_t}\right]\right| \leq \alpha \sqrt{\mathbf{x}_{u_t,a_t}^{\top}\left(\mathbf{D}_{a}^{\top} \mathbf{D}_{a}+\mathbf{I}_{d}\right)^{-1} \mathbf{x}_{u_t,a_t}},
\end{equation*}
for any $\delta>0$ and $\mathbf{x}_{u_t,a_t} \in \mathbb{R}^{d}$, where $\alpha=1+\sqrt{\ln (2 / \delta) / 2}$ is a constant. Therefore, the inequality gives a reasonably tight UCB for the expected reward of arm $a_t$, from which the arm-selection (recommendation) strategy can be derived: at each trial $t$, choose
\begin{equation*}
a_{t} \stackrel{\text { def }}{=} \arg \max _{a \in \mathcal{A}_{t}}\left(\mathbf{x}_{u_t,a_t}^{\top} \hat{\boldsymbol{\theta}}_{a}+\alpha \sqrt{\mathbf{x}_{u_t,a_t}^{\top} \left(\mathbf{D}_{a}^{\top} \mathbf{D}_{a}+\mathbf{I}_{d}\right)^{-1} \mathbf{x}_{u_t,a_t}}\right).
\end{equation*}

\medskip\noindent
Actually, the contextual bandit model improves the recommendation by leveraging the user/item features through the idea of collaborative filtering \cite{sarwar2001item,schafer2007collaborative}, i.e., those items are more likely to be recommended to a user who showed preference for items with similar features.

There are also studies pointing out that exploration in recommendations is important, i.e., the recommendations should be diverse instead of being limited by similar items \cite{qin2014contextual,liu2020diversified,ding2020hybrid}. For instance, \citet{ding2020hybrid} consider the fact that users may have different preference with regard to the diversity of items, e.g., a user with specific interest may prefer a relevant item set than a diverse item set, while another user without specific interest may prefer a diverse item set to explore his interests. Therefore, the authors propose a bandit learning framework to consider the user’s preferences on both the item relevance features and the diversity features. It is a way to trade off the accuracy and diversity of recommendation results.

Besides, \citet{10.1145/3292500.3330991} use a cascading bandit in a visual dialog augmented interactive recommender system. In cascading bandits, the user examines the recommended list from the first item to the last and selects the first attractive one \cite{kveton2015cascading,zong2016cascading}. This setting is practical to implement in online recommender systems or search engines. It has an excellent advantage as it can provide reliable negative samples, which are critical for recommendation, and the problem has drawn a lot of research attention \cite{chen2019samwalker,ijcai2019-309,wang2020reinforced,lian2020Personalized,chen2019samwalker}. Since the system can ensure that the items before the first selected one are not attractive, thus it can easily obtain reliable negative samples. Another contribution is the use of the item's visual appearance and user feedback to design more efficient exploration. 

In addition, there are other efforts to enhance bandit methods in different recommendation scenarios. For instance, \citet{chou2015pseudo} indicate that a user would only choose one or a few arms in the candidates, leaving out the informative non-selected arms. They propose the concept of pseudo-rewards, which embeds estimates to the hidden rewards of non-selected actions under the bandit setting. 
\citet{wang2018online} consider dependencies among items and explicitly formulate the item dependencies as clusters on arms, where arms within a single cluster share similar latent topics.  They adopt a generative process based on a topic model to explicitly formulate the arm dependencies as the clusters on arms, where dependent arms are assumed to be generated from the same cluster. 
\citet{yang2020hierarchical} consider the situations where there are exploration overheads, i.e., there are non-zero costs associated with executing a recommendation (arm) in the environment, and hence, the policy should be learned with a fixed exploration cost constraint. They propose a hierarchical learning structure to address the problem.
\citet{sakhi2020blob} state that the online bandit signal is sparse and uneven, so they utilize the massive offline historical data. The difficulty is that most of offline data is irrelevant to the recommendation task, and the authors propose a probabilistic model to solve it. 

\medskip\noindent%
The advantage of multi-armed bandit methods is their ability to conduct online learning, enabling the model to learn the preferences of cold users and adjust the strategy quickly after several trials to pursue a global optimum.

\subsection{Multi-Armed Bandits in CRSs}

\begin{figure}[pos=t]
\centering
\includegraphics[width=1\linewidth]{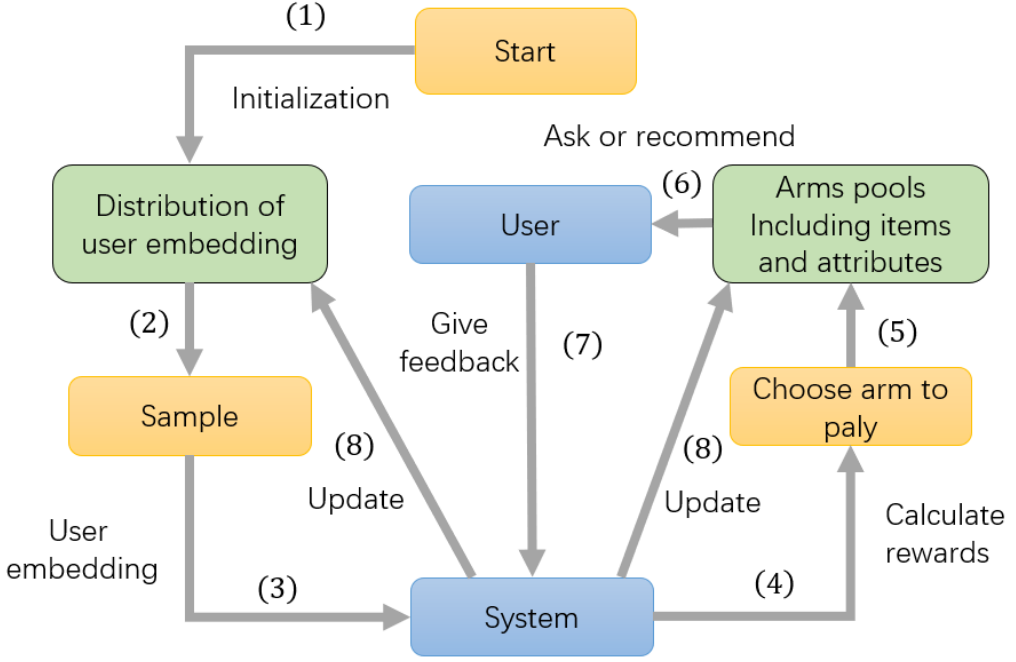}
\caption {The flowchart of the ConTS algorithm. Credits: \citet{li2020seamlessly}.}
\label{fig:conTS}
\end{figure}


The ability to interact with users enables \acp{CRS} to directly use \ac{MAB}-based methods to help the recommendation. \citet{christakopoulou2016towards} propose a classic \ac{CRS} based on \ac{MAB}, which uses several naive \ac{MAB}-based methods to enhance the offline \ac{PMF} model \cite{mnih2007probabilistic}. They first initialize the model parameters using offline data, then leverage real-time user feedback to update parameters via several common multi-armed bandit models, including the aforementioned greedy strategy, random strategy, UCB \cite{auer2002using,auer2002finite}, and TS \cite{chapelle2011empirical}. On the one hand, the performance improves on the initialized model due to the online updating; on the other hand, the offline initialization helps bandit methods reduce the computational complexity.

As mentioned above, the original \ac{MAB} methods ignore item features, which could be very helpful in recommendation. Hence, \citet{zhang2020conversational} propose a conversational upper confidence bound (ConUCB) algorithm to apply the LinUCB model \cite{li2010contextual} in the \ac{CRS} context. Instead of asking items, ConUCB asks the user about one or more attributes (key-terms in their work).
Specifically, they make the assumption that user preference on attributes can propagate to items, hence the system can analyze user feedback on queried attributes to quickly narrow down the item candidates. 
The strategies to select the attributes and arms depend on both the attribute-level and arm-level rewards, i.e., the feedback on attributes and items will be absorbed into the model parameters for future use. In addition, 
the authors employ a hand-crafted function to determine the timing to ask attributes or make recommendation, e.g., making $k$ conversations in every $m$ rounds.

However, hand-crafted strategies are fragile and inflexible, as the system should make recommendation only when the confidence is high. Therefore, \citet{li2020seamlessly} propose a Conversational Thompson Sampling method (ConTS) to automatically alternate asking questions about attributes with recommending items. They achieve this goal by unifying all attributes and items in the same arm pool, thus an arm selected from the arm pool can be either a recommendation about an item or a question about an attribute. The flowchart of ConTS is illustrated in \myfig{conTS}. ConTS assumes each user's preference vector $\tilde{\mathbf{u}}$ is sampled from a prior Gaussian distribution as $\tilde{\mathbf{u}} \sim \mathcal{N}\left(\boldsymbol{\mu}_{u}, l^{2} \mathbf{B}_{u}^{-1}\right)$, where the $\boldsymbol{\mu}_{u}$, $\mathbf{B}$, and $l$ are parameters.

\begin{table*}[pos=t]
\caption{E\&E-based methods adopted by interactive recommender systems (IRSs) and \acp{CRS}.}
\label{tab:MAB}
\tabcolsep=20pt
\renewcommand\arraystretch{1.3}
\begin{tabular}{@{}lll@{}}
\toprule
                                      & \textbf{Mechanism}                                    & \textbf{Publications}                                       \\ \midrule
\multirow{7}{*}{\textbf{MAB in IRSs}} & Linear UCB considering item features                  & \cite{li2010contextual}                                    \\
                                      & Considering diversity of recommendation               & \cite{qin2014contextual,liu2020diversified,ding2020hybrid} \\
                                      & Cascading bandits providing reliable negative samples & \cite{kveton2015cascading,zong2016cascading}               \\
                                      & Leveraging social information	                      & \cite{10.1145/3292500.3330991}                             \\ 
                                      & Combining offline data and online bandit signals      & \cite{sakhi2020blob}                                       \\
                                      & Considering pseudo-rewards for arms without feedback  & \cite{chou2015pseudo}                                      \\
                                      & Considering dependency among arms                     & \cite{wang2018online}                                      \\
                                      & Considering exploration overheads                     & \cite{yang2020hierarchical}                                \\ \midrule 
\multirow{4}{*}{\textbf{MAB in CRSs}} & Traditional bandit methods in CRSs                    & \cite{christakopoulou2016towards}                          \\
                                      & Conversational upper confidence bound                 & \cite{zhang2020conversational}                             \\
                                      & Conversational Thompson Sampling                      & \cite{li2020seamlessly}                                    \\ 
                                      & Cascading bandits augmented by visual dialogues       & \cite{10.1145/3292500.3330991}                             \\ \midrule
\textbf{Meta learning for CRSs}       & Learning to learn the recommendation model        & \cite{lee2019melu,zou2020neural,wei2020fast}               \\ 
\bottomrule 
\end{tabular}
\end{table*}

For each new-coming user, the mean of prior Gaussian distribution, $\boldsymbol{\mu}_{u}$, is initialized by the average of existing users' preference vector $\mathcal{U}^{\text {old}}$ as:
\begin{equation}
\boldsymbol{\mu}_{u}=\frac{1}{|\mathcal{U}^{\text {old}}|} \sum_{i=1}^{|\mathcal{U}^{\text {old}}|} \mathbf{u}_{i}, \mathbf{u}_{i} \in \mathcal{U}^{\text {old }}.
\end{equation}
The expected reward of arm $a$ (which can either be an item or an attribute) for user $u$ is also formulated as a Gaussian distribution since the Gaussian family is conjugate to itself. The expected reward is written as:
\begin{equation}
r\left(a, u, \mathcal{P}_{u}\right) \sim \mathcal{N}\left(\tilde{\mathbf{u}}^{\top} \mathbf{x}_{a}+\sum_{p_{i} \in \mathcal{P}_{u}} \mathbf{x}_{a}^{\top} \mathbf{p}_{i}, l^{2}\right),
\end{equation}
where $\mathcal{P}_{u}$ denotes the user's currently known preferred attributes obtained in historical conversations. And $\mathbf{x}_a$ represents the embedding vector of an arm. In the reward function, the term $\tilde{\mathbf{u}}^{\top} \mathbf{x}_{a}$ models the general preference of user $u$ to arm $a$, and the term $\sum_{p_{i} \in \mathcal{P}_{u}} \mathbf{x}_{a}^{\top} \mathbf{p}_{i}$ models the affinity between arm $a$ and the user’s preferred attributes $\mathcal{P}_{u}$. Then ConTS select an arm with the maximal reward as:
\begin{equation}
a(t)=\operatorname{argmax}_{a \subset \mathcal{A}_{u}} \tilde{\mathbf{u}}^{\top} \mathbf{x}_{a}+\sum_{p_{i} \in \mathcal{P}_{u}} \mathbf{x}_{a}^{\top} \mathbf{p}_{i}.
\end{equation}
Note that if the $a(t)$ is an attribute, the system will query the user about the preference on this attribute; if it is an item, the system will make a recommendation using this item. After obtaining users' feedback, parameters such as $\boldsymbol{\mu}_{u}, \mathcal{P}_{u}, \mathbf{B}$ will be updated accordingly.

\subsection{Meta Learning for CRSs}

Beyond multi-armed bandits, there are work trying to balance between exploration and exploitation via meta learning. For instance, \citet{zou2020neural} formulate the interactive recommendation as a meta-learning problem, where the objective is to learn a learning algorithm that takes the user’s historical interactions as the input and outputs a model (policy function) that can be applied to new users. The authors follow the idea of meta reinforcement learning \cite{duan2016rl} and use Q-Learning \cite{mnih2013playing} to learn the recommendation policy. The exploration strategy is the aforementioned $\epsilon$-greedy, where the model will select the items of maximum Q-value with probability $1-\epsilon$, and choose random items with probability $\epsilon$. 

In addition, \citet{lee2019melu} address the cold-start problem in recommendation via a model based on the Model-Agnostic Meta-Learning (MAML) algorithm \cite{pmlr-v70-finn17a}. The learned recommendation model can quickly adapt to the cold user preference in the fine-tuning stage by asking the cold user a few questions about certain items (called the evidence candidates in the work). 
A drawback of this work is that the evidence candidates are only selected once, and the query process is conducted only at the beginning when cold users arrived. It could be better to extend this strategy to a \ac{CRS} setting and develop a dynamic multi-round query strategy to further enhance the recommendation.

\subsection{Section Summary}
In this section, we introduce how a \ac{CRS} can solve the cold-start problem and trade off the E\&E balance via the interactive models such as MAB-based methods and meta learning methods. The solutions are summarized in \mytable{MAB}.
It still has a lot of room for \acp{CRS} to develop potential models to address the E\&E problem, in order to improve the user experience.

\section{Evaluation and User Simulation}
\label{sec:evaluation}
In this section, we discuss how to evaluate \acp{CRS}, which is an underexplored problem. We group attempts to evaluate \acp{CRS} into two classes: 
\begin{enumerate*}[label=(\arabic*)]
\item Turn-level evaluation, which evaluates a single turn of the system output, including the recommendation task and response generation task, which are both supervised prediction tasks.
\item Conversation-level evaluation, which evaluates the performance of the multi-turn conversation strategy which is a sequential decision making task. To achieve the goal, user simulation is important.
\end{enumerate*}
We first introduce the commonly used datasets in \acp{CRS}, and then we introduce the metrics, methods, and problems in the turn-level and conversation-level evaluation of \acp{CRS}. Finally, we discuss the strategies of user simulation in \acp{CRS}.

\begin{table*}[pos=t]
\tabcolsep=5pt
\scriptsize
\renewcommand\arraystretch{1.7}
\caption{Statistics of commonly used datasets of \acp{CRS}.}
\label{tab:dataset}

\begin{tabular}{lllllll}
\hlinew{1pt}
\textbf{Dataset}                                        & \textbf{\#Dialogs}                   & \textbf{\#Turns}                   & \textbf{Dialogue Type}                  & \textbf{Domains}               & \textbf{Dialogue Resource}                                                                             & \textbf{Related Publications}                                                           \\ \hlinew{1pt}
\textbf{MovieLens}~\cite{Bertin-Mahieux2011}                                                       & \multicolumn{3}{c}{\multirow{4}{*}{\begin{tabular}[c]{@{}c@{}}Depended on the dialogue\\ simulation process\end{tabular}}} & Movie                          & From item ratings                                                                & \begin{tabular}[c]{@{}l@{}}\cite{cikm13/ICF,loepp2014choice,vendrov2020gradient,zou2020neural}, \\ \cite{lee2019melu,iovine2020conversational,IAI2020cikm}\end{tabular}\\
\textbf{LastFM}~\cite{Bertin-Mahieux2011}                               & \multicolumn{3}{c}{}                                                                                                       & Music                          & From item ratings                                                                       & \cite{lei20estimation,lei2020interactive,zhou2020leveraging}                             \\
\textbf{Yelp}                                                           & \multicolumn{3}{c}{}                                                                                                       & Restaurant                     & From item ratings                                                                                  & \cite{Sun:2018:CRS:3209978.3210002,lei20estimation,lei2020interactive}          \\
\textbf{Amazon} \cite{mcauley2015image}                 & \multicolumn{3}{c}{}                                                                                                       & E-commerce                     & From item ratings                                                                                      & \begin{tabular}[c]{@{}l@{}}\cite{zhang2018towards,fu2020cookie,10.1145/3397271.3401180,bertCRS2020recysc},\\\cite{10.1145/3298689.3347009,luo2020deep,luo2020latent,HOOPS}\end{tabular}                    \\ \hline
\textbf{TG-ReDial} \cite{zhou2020topicguided}           & 10,000                                 & 129,392                                 & Rec., chichat                           & Movie, multi topics            & \begin{tabular}[c]{@{}l@{}}From item rating, and\\ enhanced by multi topics\end{tabular} & \cite{zhou2020topicguided}                                                      \\
\textbf{Facebook\_Rec} \cite{dodge2015evaluating}       & 1M                                     & 6M                                      & Rec.                                    & Movie                          & From item ratings                                                                                      & \cite{dodge2015evaluating}                                                      \\
\textbf{COOKIE} \cite{fu2020cookie}                     & Not given                                   & 11,638,418                              & Rec.                                    & E-commerce                  & \begin{tabular}[c]{@{}l@{}}From interactions and reviews \\ on Amazon dataset \cite{mcauley2015image}\end{tabular}                      & \cite{fu2020cookie}                                                             \\
\textbf{HOOPS} \cite{HOOPS} & Not given                                  & 11,638,418                                 & Rec.                                    & E-commerce                          & \begin{tabular}[c]{@{}l@{}}From interactions and reviews \\ on Amazon dataset \cite{mcauley2015image}\end{tabular}                                                                                   & \cite{HOOPS} \\
\textbf{DuRecDial} \cite{liu2020ACL}                    & 10,190                                 & 155,477                                 & Rec., QA, etc.                          & Movie, restaurant, etc.        & Generated by workers                                                                                   & \cite{liu2020ACL}                                                               \\
\textbf{OpenDialKG} \cite{moon-2019-opendialkg}         & 15,673                                 & 91,209                                  & Rec. chitchat                           & Movie, book, sport, etc.      & Generated by workers                                                                                   & \cite{moon-2019-opendialkg}                                                     \\
\textbf{ReDial} \cite{nips18/DeepConv}                  & 10,006                                 & 182,150                                 & Rec., chitchat                          & Movie                          & Generated by workers                                                                                   & \cite{nips18/DeepConv,chen-etal-2019-towards,zhou2020improving,ma2020bridging}                \\
\textbf{MGConvRex} \cite{xu2020user}                    & 7.6K+                                  & 73K                                     & Rec.                                    & Restaurant                     & Generated by workers                                                                                   & \cite{xu2020user}                                                               \\
\textbf{GoRecDial} \cite{kang-etal-2019-recommendation,ma2020bridging} & 9,125                                  & 170,904                                 & Rec.                                    & Movie                          & Generated by workers                                                                                   & \cite{kang-etal-2019-recommendation}                                            \\ 
\textbf{INSPIRED} \cite{emnlp20_inspired} & 1,001                                  & 35,811                                 & Rec.                                    & Movie                          & Generated by workers                                                                                   & \cite{emnlp20_inspired} \\                                            
\textbf{ConveRSE} \cite{iovine2019dataset} & Not given                                  & 9,276                                 & Rec.                                    & Movie, books, music                          & Generated by workers                                                                                   & \cite{iovine2019dataset,iovine2020conversational} \\ 
\hlinew{1pt}
\end{tabular}
\end{table*}

\subsection{Datasets and Tools}
\label{sec:data-tools}

We list the statistics of the commonly used \ac{CRS} datasets in \mytable{dataset}. Some studies collect human-human and human-machine conversation data by asking true users to converse using natural language under certain rules. To guarantee the quality of the data, these users will be rewarded after providing qualified data. There are crowdsourcing sites, such as \acf{AMT}\footnote{\url{https://www.mturk.com/}}, where the researchers can find participants to fulfill their data collection task \cite{nips18/DeepConv,moon-2019-opendialkg,liu2020ACL,emnlp20_inspired}. 

As mentioned earlier, a lot of studies of \ac{CRS} focus on the interaction policy and the recommendation strategy instead of language understanding and generation. Thus, all these studies need is the labeled entities (including users, items, attributes, etc.) in the multi-turn conversation \cite{zhang2018towards,christakopoulou2018q,lei20estimation,lei2020interactive,li2020seamlessly,HOOPS}. These studies mainly simulate and construct the user interaction from the historical records in traditional recommendation datasets, e.g., MovieLens \cite{Bertin-Mahieux2011}, LastFM \cite{Bertin-Mahieux2011}, Yelp\footnote{\url{https://www.yelp.com/dataset}}, and Amazon dataset \cite{mcauley2015image}. 


Although it seems to be many datasets in \acp{CRS}, these datasets are not qualified to develop the \acp{CRS} that can work in industrial applications. The reason is twofold: first, the scale of these datasets is not enough to cover the real-world entities and concepts; second, the conversation is either constructed from the non-conversation data or generated under certain rigorous constraints, so it is hard to generalize to the complex and diverse real-world conversations.
Therefore, more effort is needed to develop large-scale, generalizable, natural datasets for \acp{CRS}.
Therefore, more effort is still needed to develop large-scale, generalizable, diverse, and natural datasets for \acp{CRS}.

There are many different settings in \acp{CRS}, making comparison between different models difficult. Recently, \citet{zhou2101crslab} have implemented an open-source toolkit, called CRSLab\footnote{\url{https://github.com/RUCAIBox/CRSLab}}, for building and evaluating \acp{CRS}. They unify the tasks in existing \acp{CRS} into three sub-tasks: namely recommendation, conversation and policy, which correspond to our three components in \myfig{framework}: recommendation engine, user interface, and conversation strategy module, respectively. Some models and metrics are implemented under the three tasks, and the toolkit contains an evaluation module that can not only 
conduct the automatic evaluation but also the human evaluation through an interaction interface, which makes the evaluation of \acp{CRS} more intuitive. However, up to now, the majority of implemented methods are based on end-to-end dialogue systems \cite{nips18/DeepConv,chen-etal-2019-towards,zhou2020improving} or deep language models \cite{zhou2020topicguided}; the \acp{CRS} that focus on the interaction policy and the multi-turn conversation strategies (\cite{lei2020interactive,lei20estimation}) are absent.

\subsection{Turn-level Evaluation}
The fine-grained evaluation of \acp{CRS} is conducted on the output of each single turn, which contains two tasks: language generation and recommendation.

\subsubsection{Evaluation of Language Generation}

For \ac{CRS} models that generate natural language-based responses to interact with users, the quality of the generated responses is critical. Thus we can adopt the metrics used in dialogue response generation to evaluate the output of \ac{CRS}. Two example metrics are BLEU \cite{papineni-etal-2002-bleu} and Rouge \cite{lin-2004-rouge}. BLEU measures the precision of generated words or n-grams compared to the ground-truth words, representing how much the words in the machine-gene\-rated utterance appeared in the ground-truth reference utterance. Rouge measures the recall of it, i.e., how many of the words or n-grams in the ground-truth reference utterance appear in the machine-gene\-rated utterance. 

However, it is widely debated whether these metrics are suitable for evaluating language generation \cite{liu-etal-2016-evaluate,novikova-etal-2017-need}.  Because those metrics are only sensitive to lexical variation, they cannot appropriately assess semantic or syntactic variations of a given reference. Meanwhile, the goal of the proposed system is not to predict the highest probability response, but rather the long-term success of the dialogue. 
Thus, other metrics reflecting user satisfaction are more suitable in evaluation, such as measuring fluency \cite{celikyilmaz2018deep,narayan2018ranking,du2017learning}, consistency \cite{gandhe2008evaluation,lapata2015coherence}, readability \cite{lapata2003probabilistic}, informativeness \cite{huang2018finish}, diversity \cite{li-etal-2016-deep,ippolito-etal-2019-comparison,gao-etal-2019-jointly}, and empathy \cite{NEURIPS2019_fc981212,sharma2021towards}. For more metrics and evaluation methods on text generation, we refer the readers to the overviews \cite{celikyilmaz2020evaluation,deriu2021survey}.

However, the \acp{CRS} based on end-to-end dialogue frameworks or deep language models may have limitations regarding the usability in practice. Recently, \citet{jannachend2020} conducted an evaluation on the two state-of-the-art end-to-end frameworks \cite{nips18/DeepConv,chen-etal-2019-towards}, and showed that both models face three critical issues: 
\begin{enumerate*}[label=(\arabic*)]
\item For each system, about one-third of the system utterances are not meaningful in the given context and would probably lead to a breakdown of the conversation in a human evaluation. 
\item Less than two-thirds of the recommendations were considered to be meaningful in a human evaluation. 
\item Neither of the two systems ``generated'' utterances, as almost all system responses were already present in the training data.
\end{enumerate*}
\citet{jannachend2020}'s analysis shows that human assessment and expert analysis are necessary for evaluating \ac{CRS} models as there is no perfect metric to evaluate all aspects of a \ac{CRS}. The CRS models and their evaluation still have a long way to go.

\subsubsection{Evaluation of Recommendation}

The performance of recommendation models is evaluated by comparing the predicted results with the records in the test set. There are two kinds of metrics in measuring the performance of recommender systems:

\begin{itemize}[nosep,leftmargin=*]
    \item \textbf{Rating-based Metrics.} These metrics assume the user feedback is an explicit rating score, e.g., an integer in the range of one to five. Therefore, we can measure the divergence between the predicted scores of models and the ground-truth scores given by users in the test set. Conventional rating-based metrics include Mean Squared Error (MSE) and Root Mean Squared Error (RMSE), where RMSE is the square root of the MSE.

    \item \textbf{Ranking-based Metrics.} These metrics are more frequently used than rating-based metrics. Ranking-based metrics require that the relative order of predicted items should be consistent with the order of items in the test set. Thereby, there is no need for explicit rating scores from users, and the implicit interactions (e.g., clicks, plays) can be used to evaluate models. For example, a good evaluation result means that the model should only recommend the items in the test set, or it means that the items with higher scores in the test set should be recommended at higher ranks than the items with lower scores. Frequently used ranking-based metrics include hits, precision, recall, F1-score, Mean Reciprocal Rank (MRR), Mean Average Precision (MAP), and Normalized Discount\-ed Cumulative Gain (NDCG) \cite{NDCG}.

    Recently, it has become common for researchers to speed up evaluation by sampling a small set of irrelevant items and calculate the ranking-based metrics only on the small set \cite{he2017ncf,sigir/EbesuSF18,10.1145/3219819.3219965,Yang_2018_unbiased_evaluation}. However, \citet{10.1145/3394486.3403226} point out and prove that some metrics, such as average precision, recall, and NDCG, are inconsistent with the exact metrics when they are calculated on the sampled set. This means that if a recommender A outperforms a recommender B on the sampled metric, it does not imply that A has a better metric than B when the metric is computed exactly. Therefore, the authors suggest that sampling during evaluation should be avoided; if it is necessary to sample, using the corrected metrics proposed by the authors is a better choice.
\end{itemize}

\medskip\noindent%
The biggest problem in these evaluation methods is that real-world user interactions are very sparse, and a large fraction of items never have a chance of being consumed by a user. However, this does not mean that the user does not like any of them. Perhaps the user has never seen them, or the user just does not have resources to consume them \cite{10.1145/3397271.3401083,chen2020bias}. Hence, taking the consumed items in the test set as the users' ground-truth preferences can introduce evaluation biases \cite{Yang_2018_unbiased_evaluation,chen2020bias}. Unlike static recommender systems, \acp{CRS} have the ability to ask real-time questions, so the system can make sure whether a user is satisfied with an item by collecting users' online feedback. This online user test can avoid biases and provide conversation-level assessments for the \ac{CRS} model.

\subsection{Conversation-level Evaluation}
\label{sec:conversation-level-evaluation}

Different from the turn-level evaluation which compares the prediction results with the ground-truth labels in a supervised way, the conversation-level evaluation is not a supervised prediction task. The interaction process is not i.i.d. (independent and identically distributed) since each observation is part of a sequential process and each action the system makes can influence future observations. Plus, the conversation heavily relies on the user feedback. Therefore, the evaluation of the conversation requires either an online user test or leveraging historical interaction data which can be conducted by the off-policy evaluation or using user simulation.

\subsubsection{Online User Test}
The online user test, or A/B test, can directly evaluate the conversation policy by leveraging true user feedback. To conduct the assessment, the appropriate metrics should be designed. For example, the average turn (AT) is a global metric to optimize in a \ac{CRS}, as the model should capture user intention and make successful recommendations thus finish the conversation with as few turns as possible \cite{lei20estimation,lei2020interactive,li2020seamlessly}. A similar metric is the recommendation success rate (SR$@t$), which measures how many conversations have ended with the successful recommendation by the $t$-th turn.  Besides, the ratio of failed attempts, e.g., how many of the questions asked by the system are rejected or ignored by users, can be a feasible way to measure whether a system makes decisions to the users' satisfaction. 

Besides these global statistics, the cumulative performance of each turn of the conversation can also reflect the overall quality of the conversation. The expectation of the cumulative reward of a conversation policy can be written as:
\begin{equation}
J(\pi)=\mathbb{E}_{\tau \sim \pi(\tau)}\left[\sum_{t=0}^{T} \gamma^{t} r\left(\mathbf{s}_{t}, \mathbf{a}_{t}\right)\right],
\label{eq:online_eval}
\end{equation}
where the conversation trajectory $\tau$ is a sequence of states and actions of length $T$, $\pi(\tau)$ is the trajectory distribution under policy $\pi$. $\gamma \in(0,1]$ is a scalar discount factor. $r\left(\mathbf{s}_{t}, \mathbf{a}_{t}\right)$ is the immediate reward obtained by performing action $\mathbf{a}_{t}$ at state $\mathbf{s}_{t}$, e.g., it can be a feedback signal that reflects user satisfaction such as user clicks or dwell time \cite{chen2019top,ie2019reinforcement}.


Though effective, the online user evaluation has critical problems: 
\begin{enumerate*}[label=(\arabic*)]
\item The online interaction between humans and \acp{CRS} is slow and usually takes weeks to collect sufficient data to make the assessment statistically significant \cite{li2015toward,gilotte2018offline,zhao2019toward}.
\item Collecting users' feedback is expensive in terms of engineering and logistic overhead \cite{jagerman-when-2019,jagerman2018opensearch,xu2015infrastructure} and may hurt user experience as the recommendation may not satisfy them \cite{schnabel2018short,li2015toward,gilotte2018offline,chen2019generative}.
\end{enumerate*}
Therefore, a natural solution is to leverage the historical interaction, where the off-policy evaluation and user simulation techniques can be used.

\subsubsection{Off-Policy Evaluation}
Off-policy evaluation, also called counterfactual reasoning or counterfactual evaluation, is designed to answer a counterfactual question: what would have happened if instead of $\pi_\beta$ we would have used $\pi_\theta$? Specifically, when we want to evaluate the current target policy $\pi_\theta$ but we only have data under a behavior policy (or logging policy) $\pi_\beta$, we can still evaluate the target policy $\pi_\theta$ by introducing the importance sampling or inverse propensity score \cite{gilotte2018offline,jagerman-when-2019,chen2019top,10.1145/3394486.3403229,levine2020offline} as:
\begin{equation}
J\left(\pi_{\theta}\right)=\mathbb{E}_{\tau \sim \pi_{\beta}(\tau)}\left[\frac{\pi_{\theta}(\tau)}{\pi_{\beta}(\tau)} \sum_{t=0}^{T} \gamma^{t} r(\mathbf{s}_t, \mathbf{a}_t)\right].
\end{equation}
It is similar to \myeqp{online_eval} except we use data logged under another policy to evaluate the target policy. Where a weight $w(\tau)=\frac{\pi_{\theta}(\tau)}{\pi_{\beta}(\tau)}$ is used to address the distribution mismatch between the two policy $\pi_\beta$ and $\pi_\theta$. Unfortunately, such an estimator suffers from high variance when $\pi_\theta$ deviates from $\pi_\beta$ a lot. The variance reduction techniques are introduced as the remedy. The common techniques include weight clipping \cite{chen2019top,Pseudo-Dyna-Q} which limits $w(\tau)$ by an upper bound, and trusted region policy optimization (TRPO) \cite{TRPO,chen2019top}.

Another intuitive method is to directly simulate user behaviors just like the online user test, where user feedback is provided by the user simulators instead of true users. It is efficient and can avoid the high variance problem. However, the challenge is that the preference of simulated users may deviate from the true users, i.e., the user simulation can avoid high variance, but it introduces bias. Therefore, creating reliable user simulators is a crucial challenge.

\subsubsection{User Simulation}

There are generally four types of strategies in simulating users: 
\begin{enumerate*}[label=(\arabic*)]
\item using the direct interaction history of users, 
\item estimating user preferences on all items, 
\item extracting from user reviews, and 
\item imitating human conversational corpora.
\end{enumerate*}

\begin{itemize}[nosep,leftmargin=*]
  \item \emph{Using Direct Interaction History of Users.} The basic idea is similar to the evaluation of traditional recommender systems, where a subset of human interaction data is set aside as the test set. If the items recommended by a \ac{CRS} are in the users' test set, then this recommendation is deem\-ed to be a successful one. As user-machine interactions are relatively rare, there is a need to generate/simulate interaction data for training and evaluation. \citet{Sun:2018:CRS:3209978.3210002} make a strong assumption that users visit restaurants after chatting with a virtual agent. Based on this assumption, they create a crowdsourcing task to use a schema-based method to collect dialogue utterances from the Yelp dataset. In total, they collect $385$ dialogues, and simulate $875,721$ dialogues based on the collected dialogues by a process called delexicalization. For instance, ``I’m looking for Mexican food in Glendale'' is converted to the template: ``I’m looking for <Category> in <City>'', then they use these templates to generate dialogues by using the rating data and the rich information on the Yelp dataset.  \citet{lei20estimation,lei2020interactive} use click data in the LastFM and Yelp datasets to simulate conversational user interactions. Given an observed user-item interaction, they treat the item as the ground truth item to seek for and its attributes as the oracle set of attributes preferred by the user in this session. First, the authors randomly choose an attribute from the oracle set as the user’s initialization to the session. The session goes into a loop of a ``model acts – simulator responses'' process, in which the simulated user will respond with ``Yes'' if the query entity is contained in the oracle set and ``No'' otherwise. Most \ac{CRS} studies adopt this simulation method because of its simplicity \citep{10.1145/3397271.3401180,christakopoulou2018q,chen2019large}. However, the sparsity problem in recommender systems still remains: only a few values in the user-item matrix are known, while most elements are missing, which forbids the simulation on these items.

  \item \emph{Estimating User Preferences on All Items.} Using direct user interactions to simulate conversations has the same drawbacks as we mentioned above, i.e., a large number of items that have not been seen by a user are treated as disliked items. To overcome this bias in the evaluation process, some research proposes to obtain the user preferences on all items in advance. Given an item and its auxiliary information, the key to simulating user interaction is to estimate or synthesize preferences for this item. For example, \citet{christakopoulou2016towards} ask $28$ participants to rate $10$ selected items, and then they can estimate the latent vectors of the $10$ users' preferences based on their matrix factorization model. By adding noise to the latent vector, they simulate $50$ new user profiles and calculate these new users' preferences on any items based on the same matrix factorization model. \citet{zhang2020conversational} propose to use ridge regression to compute user preferences based on these known rewards on historical interaction and users' features; they synthesize the user's reaction (rewards) on each item according to the computed preferences. This kind of method can theoretically simulate a complete user preference without the exposure bias. However, because the user preferences are computed or synthesized, it could deviate from real user preferences. \citet{huang-2020-keeping} analyze the phenomenon of popularity bias \cite{10.1145/2043932.2043957,10.1145/2365952.2365982} and selection bias \cite{marlin2012collaborative,10.5555/3044805.3045061,10.1145/2507157.2507160} in simulators built on logged interaction data and try to alleviate model performance degradation due to these biases; it remains to be seen to which degree generated interactions of the type described above are subject to similar bias phenomena.
  
  \item \emph{Extracting Information from User Reviews.} Besides user behavior history, many e-commerce platforms have textual review data. Unlike the consumption history, an item's review data usually explicitly mentions attributes, which can reflect the users' personalized opinions on this item. \citet{zhang2018towards} transform each textual review of part of the Amazon dataset into a question-answer sequence to simulate the conversation. For example, when a user mentioned that a blue Huawei phone with the Android system in a review of a mobile phone X, then the conversation sequence constructed from this review is (Category: mobile phone $\rightarrow$ System: Android $\rightarrow$ Color: blue $\rightarrow$ Recommendation: X). 
  \citet{zhou2020topicguided} also construct simulated interactions by leveraging user reviews. Based on a given user profile and its historical watching records, the authors construct a topic thread that consists of topics (e.g., ``family'' or ``job seeking'') extracted from reviews of these watched movies. The topic thread is organized by a rule and eventually leads to the target movie. And the synthetic conversation is fleshed out by retrieving the most related reviews under corresponding topics.

  A noteworthy problem is that the aspects mentioned in reviews may contain some drawbacks of the products, which does not aid understanding why a user has chosen a product. For example, when a user complains about the capacity of a phone of $64$ Gigabytes is not enough, and it should not be simply convert to (Storage capacity: 64 Gigabytes) for the \ac{CRS} to learn. Thus, employing sentiment analysis on the review data is necessary, and only the attribute with positive sentiment should be considered as the reason in choosing the item \cite{zhang2014explicit,zhang2014users}.

  \item \emph{Imitating Humans' Conversational Corpora.} In order to generate conversational data without biases, a feasible solution is to use real-world two-party human conversations as the training data~\citep{vakulenko-2020-analysis}. By using this type of data, a \ac{CRS} system can directly mimic human behavior by learning from a large number of real human-human conversations. For example, \citet{nips18/DeepConv} ask workers from \ac{AMT} to converse in terms of the topics on the movie recommendation. Using these conversational corpora as training data, the model can learn how to respond properly based on the sentiment analysis result. \citet{liu2020ACL} conduct a similar data collection process. Except for collecting the dialogues about the recommendation, they also collect and construct a knowledge graph and define an explicit profile for each worker who seeks recommendations. Therefore, the conversational topics can contain many non-recommendation scenarios, e.g., question answering or social chitchat, which are more common in real life. To evaluate this kind of model, besides considering whether the user likes the recommended item, we have to consider if the system responds properly and fluently. The BLEU score \cite{papineni2002bleu} is used to measure the fluency of these models mimicking human conversations \cite{budzianowski2018multiwoz,zhang2019task}. 
  
  There are also drawbacks for this kind of method. First, when collecting the human conversational corpus, two workers need to enter the task at the same time, which is a rigorous setting and thus limits the scale of the dataset. Second, designers usually have many requirements that restrict the direction of the conversation. Therefore, the generated conversation is constrained and cannot fully cover the real-world scenarios. By imitating a collected corpus, learning a conversation strategy is very sensitive to the quality of the collected data. \citet{vakulenko-2020-analysis} analyze the characteristics of different human-human corpora, e.g., in terms of initative taking, and show that there are important differences between human-human and hu\-man-machine conversations.
\end{itemize}

\begin{figure*}[pos=t,align=\centering]
\centering
\includegraphics[width=0.98\linewidth]{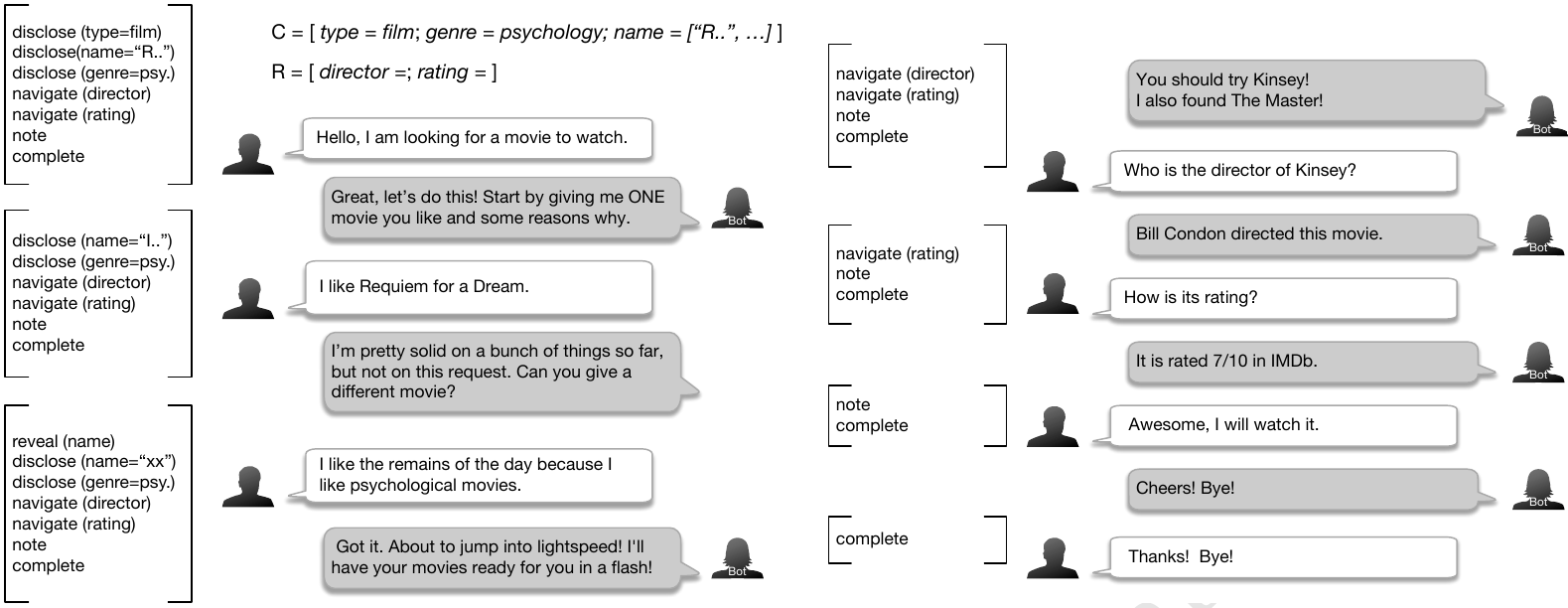}
\caption {Example dialogue with agenda sequence and state transition. The agenda is shown in square brackets. The third agenda is a result of a push operation, all other agendas updates are pull operations. Credits: \citet{simulation2020kdd}.}
\label{fig:stack}
\end{figure*}

\medskip\noindent%
Recently, \citet{simulation2020kdd} have investigated using user simulations in evaluating \acp{CRS}. They organize the action sequence of the simulated user as a stack-like structure, called the user agenda. A dynamic update of the agenda is regarded as a sequence of pull or push operations, where dialogue actions are removed from or added to the top. 
\myfig{stack} shows an example of a dialogue between the simulated user and a \ac{CRS}. 
At each turn, the simulated user updates its agenda by either a push or a pull operation based on the dialogue state and the \ac{CRS}'s action. The authors define a set of actions and the transition rule on these actions to let the simulated user imitate real users' intentions. For example, the \emph{Disclose} action indicates that the user expresses its need either actively, or in response to the agent’s question, e.g., ``I would like to arrange a holiday in Italy''. And after this action, the simulator can either transit to the \emph{Inquire} action or the \emph{Reveal} section based on how the \ac{CRS} model acts.

\medskip\noindent%
Besides modeling the user preference in simulation, another branch of studies considers modeling user behaviors in the slate, top-$K$, or list-wise recommendation. A natural solution is to consider the combinatorial action which contains a list of items instead of a single item \cite{sunehag2015deep}. However, this method is unable to scale to problems of the size encountered in large, real-world recommender systems. The feasible way is to assume a user only consumes a single item from each slate and the obtained reward only depends on the item \cite{ie2019reinforcement}. Under the assumption, user choice behavior can be modeled as the multinomial logit model \cite{louviere2000stated} or the cascade model \cite{ie2019reinforcement,10.1145/3292500.3330991,kveton2015cascading,zong2016cascading}.

Despite the recent interest in developing reliable user simulators, we believe that the research in this field is in its infancy and needs a lot of advancements.

\subsection{Section Summary}
In this section, we review the metrics, methods, and challenges in the turn-level evaluation and conversation-level evaluation of \acp{CRS}. The turn-level evaluation measures the performance of the supervised prediction tasks, i.e., recommendation and language generation of the \ac{CRS} in a single round; the conversation-level evaluation measure how the conversation strategy performs in the multi-turn conversation. Since an online user test is expensive to conduct, researchers either leverage the off-policy evaluation which assesses the target policy using the logged data under the behavior policy, or directly introduce user simulators to replace the true users in evaluation.

The evaluation of \acp{CRS} still needs a lot of effort. It ranges from constructing large-scale dense conversational recommendation data, to proposing uniform evaluation methods to compare different \ac{CRS} methods that integrate both recommendation and conversation aspects.

\section{Future Directions and Opportunities}
\label{sec:future-directions}

Having described key advances and challenges in the area \acp{CRS}, we now envision some promising future directions.

\subsection{Jointly Optimizing Three Tasks}
The recommendation task, language understanding and generation task, and conversation strategies in \acp{CRS} are usually studied separately in the three components in \myfig{framework}, respectively. The three components share certain objectives and data with each other \cite{chen-etal-2019-towards,ma2020bridging,lei20estimation,zhou2020improving}. For example, the user interface feeds extracted aspect-value pairs to the recommendation engine, and then integrates the entities produced by the recommendation engine into the generated response. However, they have the exclusive data that does not benefit each other. For instance, the user interface may use the rich semantic information in reviews but not shares with a recommendation engine \cite{nips18/DeepConv}. 
Besides, the two components may work in the end-to-end framework that lacks an explicit conversation strategy to coordinate them in the multi-turn conversation \cite{nips18/DeepConv,chen-etal-2019-towards}, thus the performance is not satisfied in human evaluation \cite{jannachend2020}.

Thereby, the three tasks should be jointly learned and guided by an explicit conversation strategy for their mutual benefit, for instance, what if the conversation strategy module were able to plan future dialogue acts based on item-item relationships (such as complementarity and substitutability \cite{mcauley2015inferring,wan2018representing,10.1145/3340531.3412332})?

\subsection{Bias and Debiasing}
It is inevitable that a recommender system could encounter various types of bias \cite{chen2020bias}. Some types of biases, e.g., popularity bias \cite{abdollahpouri2020multi,10.1145/2043932.2043957} and conformity bias \cite{zhang2014users,liu2016areyou}, can be removed with introducing the interaction between the user and system. For example,  a static recommender may not be sure whether a user will follow the crowd and like popular items. Therefore, the popularity bias is introduced in the recommender system since popular items can have higher probability of being recommended. This, however, could be avoided in \acp{CRS} because a \ac{CRS} can query about the user's attitude towards popular items in real time and avoid recommending them if the user gives negative feedback.

Nevertheless, some types of bias persist. For example, even though a recommender system may provide access to a large number of items, a user can only interact with a small set of them. If these items are chosen by a model or a certain exposure mechanism, users have no choices but to keep consuming these items. That is the exposure bias \cite{10.1145/3397271.3401083}. Moreover, users often select or consume their liked items and ignore these disliked ones even these items have been exposed to users, which introduces the selection bias \cite{marlin2012collaborative,10.5555/3044805.3045061,10.1145/2507157.2507160}, also known as the positivity bias \cite{huang-2020-keeping,10.1145/2365952.2365982}, i.e., rating data is often missing not at random and the missing ones are more likely to be disliked by the user \cite{10.5555/3044805.3045061}. These types of bias can be amplified in the feedback loop and may hurt the recommendation model \cite{sinha2016deconvolving,sun2019debiasing}. For instance, a \ac{CRS} model polluted by biased data might repeatedly generate the same items even through users suggested they would like other ones. 

There are relatively few efforts to study the bias problem in \acp{CRS}. The exploration-exploitation methods introduced in \mysec{EE} can alleviate some types of bias in \acp{CRS}. And \citet{huang-2020-keeping} make an attempt to remove the positivity bias in the user simulation stage for the interactive recommendation. Moreover, \citet{chen2020bias} present a comprehensive survey of different types of bias and describe a number of debiasing methods for \acp{RS}; it provides some perspectives for debiasing \acp{CRS}.

\subsection{Sophisticated Multi-turn Conversation Strategies}
The multi-turn strategy considered in current studies of \acp{CRS} are relatively naive. For example, there is work using a hand-crafted function to determine the timing to ask attributes or make recommendation, e.g., making $k$ conversations in every $m$ rounds \cite{zhang2020conversational}. These studies based on end-to-end dialogue systems or deep neural language models are worse: they do not even have an explicit strategy to control the multi-turn conversation \cite{nips18/DeepConv,chen-etal-2019-towards}.
Besides, some strategies can be problematic in regard to handling users' negative feedback. For instance, \citet{lei20estimation} consider updating the model parameters when the user dislikes a recommended item. However, simply taking rejected items as negative samples would influence the model's judgement on the queried attributes. For example, a user's rejection of a recreation video might be due to the fact that they watched it before, and it does not mean that they dislike recreation videos. To overcome this problem, the model should consider more sophisticated strategies such as recognizing reliable negative samples \cite{chen2019samwalker,ijcai2019-309,wang2020reinforced,lian2020Personalized,chen2019samwalker} as well as disentangling user preferences on items and attributes \cite{ma2019learning,wang2020disentangled}.

We have witnessed some studies using \ac{RL} as the multi-turn conversation strategy by determining model actions such as whether to ask or recommend \cite{Sun:2018:CRS:3209978.3210002,lei20estimation,lei2020interactive}. However, there is a lot of room for improvement in designing the state, action, and reward in \ac{RL}. For instance, more sophisticated actions can be taken into consideration such as answering open-domain questions raised by users \cite{zhu2021retrieving} or chatting non-task-oriented topics for entertainment purposes \cite{wu2018deep,liu2020ACL}. Besides, more advanced and intuitive \ac{RL} technologies can be considered to avoid the difficulties, e.g., hard to train and converge, in basic RL models \cite{wang2020deep}. For example, Inverse RL (IRL) \cite{10.5555/645529.657801} can be considered to learn a proper reward function from the observed examples in certain \ac{CRS} scenarios, where there are too many user behavior patterns so the reward is hard to define. Meta-RL \cite{duan2016rl,wang2016learning-reinforcement} can be adopted in \acp{CRS}, where the interaction is sparse and various, to speed up the training process and to improve the learning efficiency for novel subsequent tasks.

\subsection{Knowledge Enrichment}
A natural idea to improve \acp{CRS} is to introduce additional knowledge. In early stages of the development of \acp{CRS}, only the recommended items themselves were considered \cite{christakopoulou2016towards}. Later, the attribute information of items was introduced to assist in modeling user preferences \cite{christakopoulou2018q}. Even more recent studies consider the rich semantic information in knowledge graphs \cite{zhou2020improving,lei2020interactive,xu2020user,moon-2019-opendialkg}. For example, to better understand concepts in a sentence such as ``I am looking for scary movies similar to \emph{Paranormal Activity (2007)}'', \citet{zhou2020improving} propose to incorporate two external knowledge graphs (KGs): one word-oriented KG providing relations (e.g., synonyms, antonyms, or co-occurrence) between words so as to comprehend the concept ``scary'' in the sentence; one item-oriented KG carrying structured facts regarding the attributes of items.

Besides knowledge graphs, multimodal data can also be integrated into the original text-based \acp{CRS} since it can enrich the interaction from new dimensions. There are some studies that exploit the visual modality, i.e., images, in dialogue systems \cite{10.1145/3292500.3330991,Liao18mm,10.1145/3331184.3331226,10.1145/3308558.3313598}. For example, \citet{10.1145/3292500.3330991} propose a visual dialog augmented \ac{CRS} model. The model will recommend a list of items in photos, and the user will give text-based comments as feedback. The image not only helps the model learn a more informative representation of entities, but also enable the system to better convey information to the user. Except for the visual modality, other modalities can benefit \acp{CRS} and could be integrated. For example, spoken natural language can convey users' emotions as well as sentiments towards certain entities \cite{pittermann2010emotion}.

\subsection{Better Evaluation and User Simulation}
The evaluation of \acp{CRS} still has a long way to go. As we introduced in \mysec{conversation-level-evaluation}, evaluating the \ac{CRS} requires real-time feedback, which is expensive in real-world situations \cite{jagerman-when-2019}. Thus, most \acp{CRS} adopt user simulation techniques to create an environment \cite{simulation2020kdd}. However, simulated users cannot fully replace human beings. How to simulate users with maximum fidelity still needs further research. Feasible directions include designing systematic simulation agenda \cite{simulation2020kdd,schatzmann-etal-2007-agenda}, building dense user interactions for reliable simulation \cite{Pseudo-Dyna-Q,chen2019generative,NEURIPS2019_e49eb652}, and modeling user choice behaviors over the slate recommendation \cite{ie2019reinforcement,10.1145/3394486.3403229,afsar2021reinforcement}.

In addition, \acp{CRS} work on different datasets and they have various assumptions and settings. Therefore, developing comprehensive evaluation metrics and procedures to assess the performance of \acp{CRS} remains an open problem. Recently, \citet{zhou2101crslab} have implemented an open-source \ac{CRS} toolkit, enabling evaluation between different \ac{CRS} models. However, their implemented models are mainly based on end-to-end dialogue systems \cite{nips18/DeepConv,chen-etal-2019-towards,zhou2020improving} or deep language models \cite{zhou2020topicguided}, the models focusing on the explicit conversation strategy \cite{lei2020interactive,lei20estimation} are absent.

\section{Conclusion}
\label{sec:conclusion}

Recommender systems are playing increasingly important role in information seeking and retrieval. 
Despite having been studied for decades, traditional recommender systems estimate user preferences only in a static manner like through historical user behaviours and profiles. It offers no opportunities to communicate with users about their preferences. This inevitably suffers from a fundamental information asymmetry problem: a system will never know precisely what a user likes (especially when his/her preference drifts frequently) and why the user likes an item. The envision of \Acfp{CRS} brings a promising solution to such problems. With the interactive ability as well as the natural language-based user interface, \acp{CRS} can dynamically get explicit user feedback using natural languages, while increasing user engagement and improving user experience. This bold vision provides great potential for the future of recommender system, hence actively contributes to the development of the next generation of information seeking techniques.

Although the build of \ac{CRS} is an emerging field, we have spotted great efforts from different perspectives. In this survey, we acknowledge those efforts, with the aim to summarize the existing studies and to provide insightful discussions. We tentatively gave a definition of the \ac{CRS} and introduced a general framework of \acp{CRS} that consists of three components: a user interface, a conversation strategy module and a recommender engine. Based on this decomposition, we distilled five existing research directions, namely:
\begin{enumerate*}[label=(\arabic*)]
\item question-based user preference elicitation;
\item multi-turn conversational recommendation strategies;
\item dialogue understanding and generation;
\item exploitation-exploration trade\-offs for cold users;
\item evaluation and user simulation.
\end{enumerate*}
For each direction, we reviewed the existing efforts and their limitation in one section, leading to the primary structure of this survey. 
Despite the progresses on the above five directions, more interesting problems remain to be explored in the field of \acp{CRS}, such as, 
\begin{enumerate*}[label=(\arabic*)]
\item joint optimization of three components;
\item bias and debiasing methods in \acp{CRS};
\item multi-turn conversational recommendation strategies;
\item multi-modal knowledge enrichment; 
\item evaluation and user simulation.
\end{enumerate*}

Our discussions above provide a comprehensive retrospect of current progress of \acp{CRS} which can serve as the basis for the further development of this field. By providing this survey, we call arm to this emerging and interesting field. We hope this survey can inspire the researchers and practitioners from both industry and academia to push the frontiers of \acp{CRS}, making the thoughts and techniques of \acp{CRS} more prevalent for the next generation of information seeking techniques.

\section*{Acknowledgments}
This work is supported by the National Natural Science Foundation of China (61972372, U19A2079) and the National Key Research and Development Program of China (2020YFB1406703, 2020AAA0106000).

All content represents the opinion of the authors, which is not necessarily shared or endorsed by their respective employers and/or sponsors.


\bibliographystyle{cas-model2-names}
\bibliography{review}

\end{document}